# Critical elements for connectivity analysis of brain networks

Jean Faber[1], Priscila C. Antoneli[1], Guillem Via[1], Noemi S. Araújo[1], Daniel J. L. L. Pinheiro[1], Esper Cavalheiro[1,2]

[1]Universidade Federal de São Paulo - UNIFESP
Escola Paulista de Medicina (EPM), Department of Neurology and Neurosurgery, São Paulo, Brazil
[2]Centro Nacional de Pesquisa em Energia e Materiais CNPEM

Corresponding author: jean.faber@unifesp.br

## 1. INTRODUCTION

In recent years, new and important perspectives were introduced in the field of neuroimaging with the emergence of the connectionist approach (Williams and Henson 2018). In this new context, it is important to know not only which brain areas are activated by a particular stimulus but, mainly, how these areas are structurally and functionally connected, distributed, and organized in relation to other areas. In addition, the arrangement of the network elements, i.e., its topology, and the dynamics they give rise to are also important. This new approach is called connectomics (Danielle S. Bassett and Sporns 2017). It brings together a series of techniques and methodologies capable of systematizing, from the different types of signals and images of the nervous system, how neuronal units to brain areas are connected. Through this approach, the different patterns of connectivity can be graphically and mathematically represented by the so-called connectomes (Olaf Sporns, Tononi, and Kötter 2005).

The *connectome* uses quantitative metrics to evaluate structural and functional information from images of neural tracts and pathways or signals from the metabolic and/or electrophysiologic activity of cell populations or brain areas. Besides, with adequate treatment of this information, it is also possible to infer causal relationships. In this way, structural and functional evaluations are complementary descriptions which, together, represent the anatomic and physiologic neural properties, establishing a new paradigm for understanding how the brain functions by looking at brain connections (Avena-Koenigsberger, Misic, and Sporns 2017; Olaf Sporns and Kötter 2004; Olaf Sporns 2002).

A connectionist approach allows us to evaluate how the anatomic organization of the brain relates to its functional dynamics and how structural or functional changes affect this relationship (Honey, Thivierge, and Sporns 2010). In order to perform a formal and quantitative analysis, Graph Theory is used in connectomics (Fornito, Zalesky, and Breakspear 2013). This method allows a systematic, consistent, and robust evaluation of the functional and structural neural networks. In this way, since the connectome incorporates all the mathematical properties of graphs, it naturally quantifies all the properties, similarities and differences among the different neural network configurations.

Here, we highlight five critical elements of a network that allows an integrative analysis, focusing mainly on a functional description. These elements include; (*i*) the properties of its nodes; (*ii*) the metrics for connectivity and coupling between nodes; (*iii*) the network topologies; (*iv*) the network dynamics and (*v*) the interconnections between different domains and scales of network representations.

The first element we must consider is the set of intrinsic properties of the nodes that comprise a network. When considering networks at the microscopic level, it can include, the type of neurons that are connected and the ways the neurons are activated (including metabolic and electrophysiological activities; (Olaf Sporns, Tononi, and Kötter 2005). At the mesoscopic level, the anatomical circuit properties matter, as well as the specific physiological signatures and activity patterns of each connected brain layer or subfield. Finally, at the macroscopic level, attention needs to be paid to all features associated with the anatomical composition (such as neuronal density, neural subfields, tracts and arrangement within each brain area).

The second network element to highlight is the type of metric used to assess the connections or couplings between nodes (Bastos and Schoffelen 2016). Different metrics can be used to evaluate the connectivity in a brain network through quantification of the statistical dependencies, or even causal interactions, between node activities. Each one may reveal a linear or nonlinear relationship or describe a directed or undirected information flow through the network edges, defined according to what is being measured and studied. Here, we will explore three nonlinear metrics, Mutual Information, Kullback–Leibler, and Granger Causality, and one linear metric, given by the Pearson Coefficient. Furthermore, we will also discuss possible ways of coupling among the main features of the electrophysiological signal, such as amplitude, frequency, and phase (Canolty and Knight 2010).

The third element to be considered is the arrangement of the network nodes and vertices, i.e., topology (Rubinov and Sporns 2010). The topology of a brain network is one of the most important aspects of its connectome. It may reveal how a particular neural activity, or even a brain area, is relevant for a specific brain state associated with a disease or stimulus, for instance, and also how a particular neural arrangement optimizes the flow of information in a complex brain network (Kaiser 2011). Furthermore, it also describes the importance of network hubs and assemblies by quantifying their interconnections and information storage (Olaf Sporns 2013b).

The fourth element of a network is related to its change over time. The dynamic activity of the brain promotes different synchronizations among different areas and subfields at different periods of time (Danielle S. Bassett and Sporns 2017). Mainly, it is of interest to evaluate temporal phase transitions of neural activity associated with a particular cognitive state during a behavioral task or associated with a neurocognitive disorder. For example, the phase transitions related to an epileptic seizure can be described by topological changes of a network over time (Van Diessen et al. 2013). Therefore, a connectome analysis can inform how and why a neural network fluctuates, repeats, reorganizes, stabilizes, or degenerates in time.

Finally, the fifth network element addressed here is related to the most intricate brain characteristic: the way different levels of information, from biomolecules to brain areas, are integrated. From the connectome perspective, this problem might be assessed by inspecting how different networks, described at different scales, can be interconnected, and how the information storage and processing at one scale level interferes with them at another (Betzel and Bassett 2017). We are still far from having an answer on this issue, but the connectome approach allows us to propose mathematical models of integration yielding an objective formulation with a possible test of its consistency.

Through a functional and effective connectivity analysis, the type of technique used to measure brain activities is fundamentally relevant since it defines the type, the scale, and the node features of a network. For instance, techniques of invasive electrophysiological recordings can register the activity of individual cells, such as action potentials and spike trains, or the activities of groups of cells, such as local field potentials (Buzsáki, Anastassiou, and Koch 2012). Noninvasive techniques, like electroencephalography (EEG), functional magnetic resonance imaging (fMRI), magnetoencephalography (MEG) and functional near-infrared spectrum (fNIRS), also allow a connectivity analysis at a large scale. In general, EEG recordings provide a description of the overall activity of the encephalon (Babiloni et al. 2009; Pizzagalli 2007). But, despite lacking anatomical and/or physiological specificities, they can help to determine how certain cognitive or pathological mental states are associated with specific network topologies.

Clinically, the connectome approach can be extremely powerful because it provides the functional and structural topologies, quantitative parameters of the brain's activity that, in general, are not accessible using only traditional brain images (E. T. Bullmore and Bassett 2011). For example, a functional/effective connectivity analysis can help to provide information about the stability or dysfunctionality of certain neural subnetworks associated with a brain disorder such as dementia, epilepsy, or Parkinson. It also allows for an evaluation of information flow in a brain area surrounded by a tumor or around an epileptic focus (Douw et al. 2010). Another example of how connectome approach can help us to understand acute diseases or comorbidities is to study the evolution of the connection patterns over time. In addition, it brings a complete new look over the neural dynamics, addressing a truly integrated brain and not only its associated parts (Baliki et al. 2008).

## 2. BRAIN NETWORKS

### 2.1. Graph representation

Network science is an interdisciplinary field that combines concepts and techniques from computational sciences, statistics, engineering, and mathematics among others (C. J. Stam and van Straaten 2012). Through these techniques, it is possible to construct graphical models that allow for a quantitative and systematic description of how the neural systems interact, organize themselves in different geometric patterns, evolve in time and stabilize to optimize the storage, flow, and processing of information. Additionally, these models allow for statistical inferences providing evaluation and visualization of the communication process among its units along time and space and with other networks in different scales. When network analyses are applied to brain circuits, they provide robust methods to forecast structural and functional brain changes associated with specific injuries or therapeutic interventions (Cornelis J. Stam 2014; E. Bullmore and Sporns 2009a). This is called Connectomics (Fornito et al. 2012).

Connectomics is a new approach that attempts to provide a solid road for the studies of connectomes, including the diversity of neural connectivity maps in different scales of time and space (Danielle S. Bassett and Sporns 2017). In a general way, connectome descriptions are based on Graph Theory (Olaf Sporns 2011).

More formally, Graph Theory is a theoretical field dedicated to the study of mathematical structures, called graphs, used to model pairwise relations between information units. As a mathematical object, a graph $G$ can be defined by the relationship between the pair of sets for vertices $V$ and edges $E$, i.e. $G = \{V, E\}$. Edges are also known as arcs or lines and can represent the mode, type, and intensity of link between pairs of vertices. The *mode* refers to the representation of the relationship between vertices, since, for instance, a graph might be displayed in 2 or 3 dimensions. The *type* refers to the direction of connections, i.e., undirected versus directed. Finally, one must consider the *intensity* that relates the strength between connections. Vertices are also known as dots or, more commonly, *nodes* (W. Huang et al. 2018; Costa et al. 2007). Typically, a graph is represented in a schematic geometric form composed of a set of nodes graphically represented by points joined by lines or curves. The latter can be directed or undirected when intended to represent the information flow, or it can be of different thicknesses when intended to represent the degree, cost, or connection probability between the nodes (Newman 2003).

Historically, Graph Theory was introduced by Euler in 1735 when he proposed a problem known as "the seven bridges of Königsberg" (Boccaletti et al. 2006). In the German city of Königsberg, now the territory of Kaliningrad, Russia, there were seven bridges arranged in a very particular way as shown in Figure 2.1. Euler asked if it would be possible to make a path passing through all of them by crossing each bridge once. By formalizing mathematically the idea of graphs, Euler showed that this solution was impossible (Amaral and Ottino 2004).

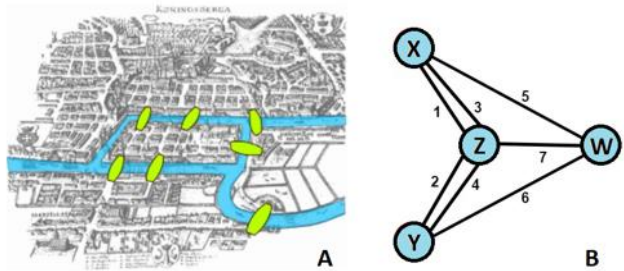

**Figure 2.1 The seven bridges of Krönigsberg.** (A) An old map of Krönigsberg with the bridge's configuration. (B) Graph representation of this map.

A graph can be expressed using matrix notations, by means of the so-called *adjacency matrices* and *incidence matrices*, that contains information about the intensity and direction of the connections between nodes. For example, given a network with N = 4 nodes, {*x, y, z, w*}, the matrix representation, and its corresponding graphs are:

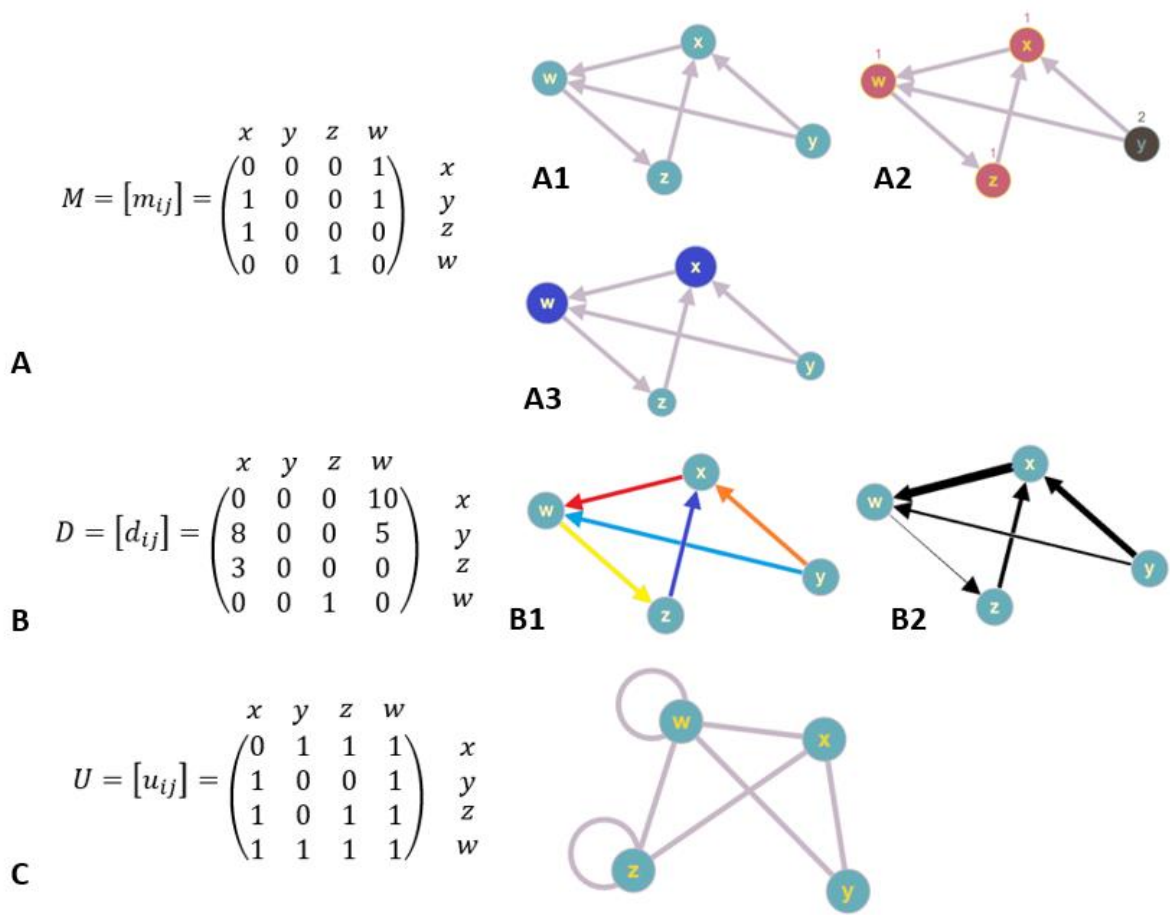

**Figure 2.2. General matrix and visual representations for different type of graphs.** (A) Matrix and visual representation for non-weighted graphs. Figure shows a visual representation of the adjacency matrix M, where each node is represented by a circle and each vertex by a directed arrow. In this representation, the vertices directions are read from row to column. (A1) Shows the number of vertices from each node for the same graph. (A2 and A3) Exhibits a graph representation highlighting the nodes with more confluences. (B) Matrix and visual representation for weighted directed graphs. Visual representation for the distance matrix D, where each node is represented by a circle and each vertex by a directed arrow. In this representation, the vertices directions are read from row to column. (B1) shows the number of vertices from each node for the same graph. (B2)

exhibit a graph representation emphasizing the nodes with more confluences. (C) Matrix and visual representation for undirected graphs with self-connections. It shows a visual representation for the symmetric matrix U, where each node is represented by a circle and each vertex by an undirected arrow (or nodes by circles and vertices by undirected arrows).

Now, considering the same nodes set {*x, y, z, w*}, we may also represent the intensity of connections by adding weights to each edge, which will be given by the values obtained from the correlation metric like those shown in Figure 2.2-B. Later, we will discuss, in more detail, different metrics and approaches to quantify statistical dependencies, correlations and couplings between nodes and their correspondent intensities.

However, for symmetrical connections given by symmetrical coupling models or metrics of statistical dependence, the edges are undirected, i.e., the links do not include arrows. These networks are represented by undirected graphs and square symmetric matrices as all the edges are bidirectional.

Graph representations enable the characterization of network patterns by evaluating how nodes are connected to each other in order to display specific topological structures. Undirected graphs, for example, are well designed for systems that incorporate symmetric coupling interactions and symmetric metrics of correlations/associations. In contrast, if one pretends to represent the flow of information between two brain areas, a directed graph describing asymmetrical connections should be considered (E. Bullmore and Sporns 2009).

In order to approach the diversity of possible connections and contact-points, graph representations can be displayed in different colors, styles, and sizes, as shown in Figure 2.2-B. The different intensities of connections can be illustrated by an index representing the strength nodes interaction, for example, using a weighted graph with different sizes of nodes and edges. It is also possible to represent graphs with multiple edges between the same pair of nodes and, also, nodes with self-connections as illustrated in Figure 2.2 - C.

The nonlinear and multiscale of brain dynamics are implicated in a diversity of physical couplings and statistical dependencies that are usually classified into three types: (*i*) functional connectivity, (*ii*) effective or causal connectivity, and (*iii*) structural connectivity (Kaiser 2011; Olaf Sporns et al. 2004).

Scans of MRI and fMRI configure the most common measurements to describe patterns of brain connections (Kaiser 2011). However, when considering evaluations of functional connectivity, other brain signals can also provide useful information to describe neural network configurations. Actually, by evaluating functional connectivity we gain a totally different perspective of brain dynamics since it allows a detailed look on how the brain uses different ways to signalize and processes information (Shine et al. 2018; Allen et al. 2018).

Currently, there are different methods for recording brain signals of different physical nature and spatiotemporal scales. Besides the fMRI technique, that essentially measures variations of blood flow associated with neural activity (Li et al. 2009), during cognitive and/or behavioral tasks (Barch et al. 2013; Rissman, Gazzaley, and D'Esposito 2004), other non-invasive techniques such as EEG, magnetoencephalography (MEG) or functional near-infrared spectrum (fNIRS) also provide information from the brain activity (Baker et al. 2018; H. Zhang et al. 2010). Since these techniques are not invasive, most of the networks yielded by these signals are related to global aspects of brain dynamics and functions. There are some computational approaches, which typically use multivariate statistics, and that allow one to infer structural signal sources from deep brain areas through EEG recordings (Grech et al. 2008). Although these approaches allow for the reconstruction of some deep brain areas and the generation of useful 3D functional network pictures, they lack the spatial accuracy of fMRI scans because they are statistical estimations (Bradley et al. 2016; Pascual-Marqui 1999). Other invasive techniques are also used to assess the electrophysiological activity of deep brain regions, such as the hippocampus, thalamus, or cerebellum, among others. This is usually done during (pre) surgical procedures, chronically implanted in patients with deep brain stimulators (DBS), animal models or voluntary subjects from brain-computer interface projects (Chen et al. 2012; Engel et al. 2005; Lal et al. 2005).

To calculate adjacency/incidence matrices reflecting functional or effective connectivity, pairs of time series are "linked" by a mathematical metric able to capture a statistical dependence (causal or not) between those sources of brain activity. In this way, the network edges in a functional connectivity description are labeled by numerical values measured by a specific statistical dependence metric that expresses a linear or nonlinear, symmetric or asymmetric correlation/association; and each network node represents the source of a measured brain signal.

## 2.2. Brain network nodes

The specification of a node depends on what we want to know in order to select and measure a specific biological/physical feature that will be used to perform the quantitative analysis. Consider, for instance, a culture of neurons; we can construct a structural connectivity network by considering single neurons as nodes and all related structural connections among other neurons through axons and dendrites as the edges. To measure the nodes and edges in this network, we can apply biomolecular and histological techniques to mark and identify all the neurons and their respective physical connections (O. Sporns 2016). Considering another scale, different brain regions can be considered as nodes in a structural representation. These regions can be the visual cortex, thalamus, hippocampus, motor cortex, etc., and the edges can be the neural pathways such as nerve fibers and tracts. In this case, the best approach to measure them is tractography which comprises a 3D imaging modeling technique that represents neural tracts by using diffusion-weighted images (DWI) recorded from MRI in parallel to computer-based image analysis (Craddock et al. 2013).

However, to evaluate and construct functional networks the possibilities are even higher. A functional connectivity analysis will be based on the type of feature or signal being measured from different neural structures. Figure 2.3 provides a short list of the main biological structures considered as nodes in structural network analysis and their possible corresponding features to be measured by means of different techniques.

| FRAMEWORK OF POSSIBLE BRAIN NODES | | |
|---|---|---|
| STRUCTURAL NODES | SIGNALIZATION : FUNCTIONAL NODES | MEASUREMENT TECHNIQUES |
| 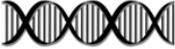 | GENE EXPRESSION | NORTHERN BLOT, REVERSE TRANSCRIPTION POLYMERASE CHAIN REACTION, DNA MICROARRAY, RNA-SEQUENCING |
| 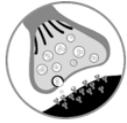 | BIOMOLECULAR CELL SIGNILING INTRACELLULAR POTENTIALS NEUROTRANSMITIONS IONIC CONCENTRATION | FLUORESCENCE CORRELATION SPECTROSCOPY, SHARP ELECTRODE, IMMUNOFLUORESCENCE, ION SELECTIVE ELECTRODE |
| 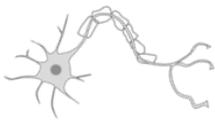 | BIOMOLECULAR SIGNILING MEMBRANE AND INTRA/EXTRACELLULAR POTENTIALS FLUORECENCE CELL IMAGING | SINGLE UNITY RECORDING, PATCH CLAMP, VOLTAGE AND CURRENT CLAMP, FLUORECENCE IMAGING |
| 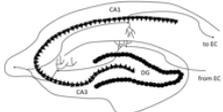 | BIOMOLECULAR SIGNILING INTRA/EXTRACELLULAR POTENTIALS FLUORECENCE CELL EXPRESSION METABOLIC CELL EXPRESSION | MULTIUNITY RECORDING, MICROELETRODES, FLUORECENCE IMAGING, IMUNOHISTOCHEMISTRY |
| 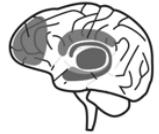 | INTRA/EXTRACELLULAR POTENTIALS ELECTRIC AND MAGNETIC BIOPOTENTIALS METABOLIC EXPRESSION | MULTIUNITY RECORDING, INVASIVE PROBES, EEG, ECoG, fNIRS, fMRI, PET, MEG |

**Figure 2.3 General framework for different type of structural and functional nodes.** The first column of the table presents a series of different possible structural nodes, according to its biological nature, information, and spatial scale. From the top to down it is listed as a possible network node: (I) a gene, (II) biomolecular, membrane and synapses, (III) cell unity (neurons and astrocytes), (IV) subfields or layers of a specific brain region, such as CA1, CA2, CA3, and CA4 of hippocampus, and (V) brain regions, such as motor cortex, visual cortex, thalamus and etc. The second column lists the main biological features evaluated in each correspondent structural node. These features are chosen according to the scientific field of investigation and each one can represent a functional node. The third column lists the main techniques used to measure the associated features of each structural node.

In summary, there are many different brain (bio)physical quantities that can be measured with a technique that reads one or more features and these measures can be characterized and compared using a wide variety of metrics. For instance, considering again a culture of cells or sliced brain tissues recorded in an *in vitro* procedure. We can record the extracellular potentials of the electrophysiological activity from particular populations of neurons or from a specific brain subfield (Poli, D., Pastore, V. P. Massobrio 2015). Therefore, the individual neural activity recorded from each electrode during a right period can be considered a node in a functional network and the edges, some possible statistical dependency among them. As it will be described later, the statistical dependencies will be totally determined by a mathematical metric and the signal feature (phase, frequency, amplitude, time, etc.) being considered (Figure 2.4).

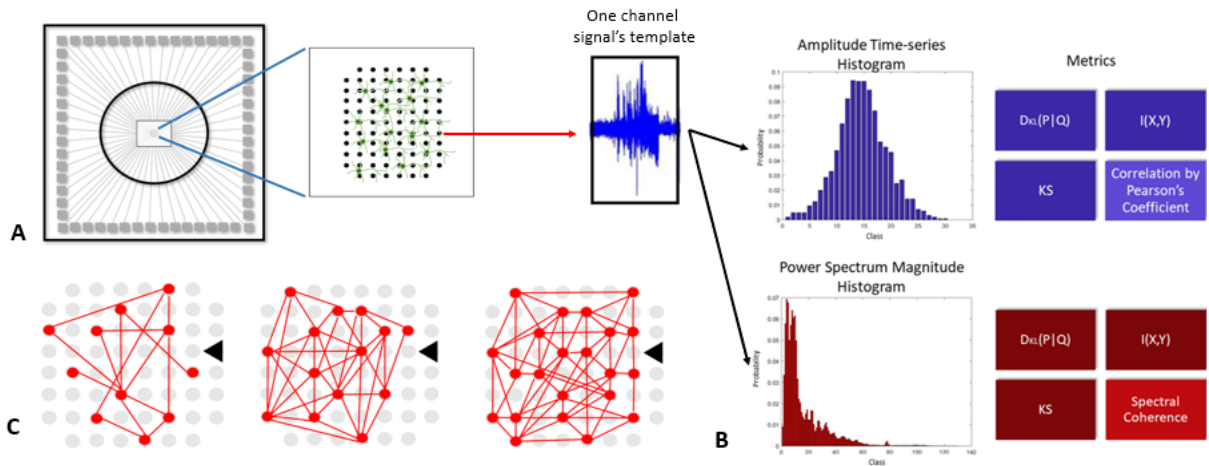

**Figure 2.4. Different kind of nodes and correlation metrics in structural and functional networks.** (A) In vitro culture of neurons on a multielectrode array (MEA) to record electrophysiological extracellular activity and two examples of two different topological configurations for a functional connectivity analysis. (B) A signal template cut from an electrophysiological recording is selected and two statistical features were calculated, amplitude histogram and power spectral density. Each one of these features, and any other, can be interpreted as a functional node of a network. (C) By means of different mathematical metrics (linear and nonlinear), it is possible to establish how all nodes are correlated. It is shown four possible metrics: $D_{KL}(P|Q)$ – Kullback–Leibler or Divergent Entropy, $I(X,Y)$ – Mutual Information, KS – Kolmogorov-Smirnov and $\rho_{XY}$ – Pearson's Coefficient.

Under the Graph Theory point of view, a node is a redistribution point or a communication endpoint. A node is defined as an *active system* attached to a network, capable of creating, receiving, or transmitting information over a communication channel known as an edge in this case (Newman 2003). Any passive distribution point, such as a *distribution frame* or *patch panel* in computer networks, is consequently not a node.

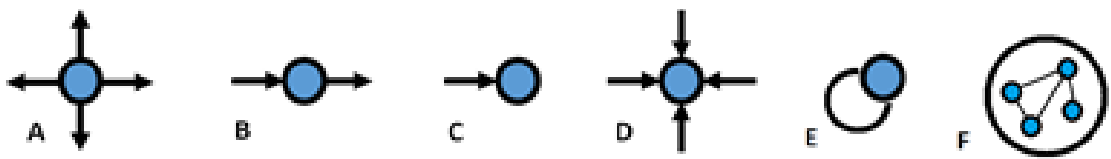

**Figure 2.5. General scheme of types of nodes.** (A) A source or client node where all point out from the node to other nodes. This type of node represents the creation of information. (B) A peer or repeater node, (C) a terminal or isolate node, (D) a sink or client node and (E) a self-reference node and (F) modular nodes.

A node can be also be classified according to their trespassed information flow as a place in a network where a message can be created (called source or server), received (sink or client), or transmitted (repeaters or peers), see Figure 2.5. A peer node sometimes may work as a client- or a server-node (Yoneki, Hui, and Crowcroft 2008). Furthermore, a peer-to-peer node or an overlay network that actively routes data to other network structures in a different spatial or temporal scale can be called *supernodes* (Navlakha, Rastogi, and Shrivastava 2008). Others information associated with network nodes are listed in Table 2.1.

**Table 2.1 Node Information.**

| | |
|---|---|
| *degree* | Number of edges associated with a node |
| *nearest* | Nearest neighbors within radius |
| *indegree* | In-degree: number of oriented edges entering the node. |
| *outdegree* | Out-degree: number of oriented edges incoming the node. |
| *predecessors* | Number of predecessors in an oriented network |
| *successors* | Number of successors in an oriented network |

To characterize the entire network, it is crucial to know the characteristics of a node in a network. It helps to define its topology which determines its efficiency in transmitting and storing information (E. Bullmore and Sporns 2009).

An important concept is the modular nodes (Figure 2.5 - F) characterized by clusters of nodes or nodes more densely connected to other nodes (Rubinov and Sporns 2010). These special nodes in neural circuits or structural networks may represent the nucleus with a specialized function, where different modules may work in parallel to support different neurophysiological processes (Olaf Sporns 2013b). In addition, a network having a distributed "core-to-periphery" configuration has a set of central nodes that are interconnected with all other nodes in the network and a set of nodes on the periphery that are sparsely interconnected with all other nodes in the network (Rubinov and Sporns 2010). This type of network architecture represents a process of information integration through neural assemblies, neural circuits, or functional nuclei, characterizing a control point (Park and Friston 2013; Olaf Sporns 2002).

All these characterizations of edges and nodes are critical and configure the first elements of a network and provide the main base to any connectivity analysis.

## 3.  MEASURES OF CONNECTIVITY

### 3.1.  Couplings and correlations

After defining the graph nodes and what they represent, it is necessary to quantify the interactions between them. These interactions are measures that denote the information provided by the graph, such as its topology, architecture, and complexity. Essentially, the edges indicate the nodes that are currently linked and the strength of this link when the graph is weighted. Thus, a graph can be symmetrical or not depending on the metrics used to calculate its connectivity.

Besides structural connectivity, one can also consider functional and effective connectivity. Functional connections refer to any form of statistical dependence between the activities of two nodes (Friston 2011), in general, without any assumption of causal influences. A statistical dependence between two variables can be defined in terms of Bayes's rule. This rule states that two variables X and Y are dependent when, at least, the probability to get one of the outcomes from either of the two variables depends on whether we conditioned it to some knowledge on obtained outcomes from the other (Altman and Krzywinski 2015). Mathematically, this can be expressed as $P(X|Y) \neq P(X)$, where $P(X|Y) = P(X,Y)/P(Y)$ refers to the probability distribution of X when its measure is conditioned to some knowledge from the outcomes of Y, and P(X,Y) is the joint probability distribution of X and Y. Inversely, X and Y are independent when P(X|Y) = P(X). The same statements apply if one exchanges X with Y in these expressions. This rule represents the more general form to measure a relationship between two random variables. Correlations are special cases of statistical dependencies, where we consider them as a mathematical metric that measures an increasing or a decreasing trend, linear or nonlinear, parametric or non-parametric. We say two variables *X* and *Y* are correlated with increasing trend when the values of *Y* increase according to the positive increase of *X* (Altman and Krzywinski 2015). Similarly, a correlation with a decreasing trend occurs when the values of *Y* decrease according to the positive increase of X.

Although a correlation metric can be linear or nonlinear, linear metrics are more commonly used in the literature, like the Pearson's correlation coefficient. In this way, functional connectivity evaluates only statistical dependencies of node outputs (Figure 3.1-A).

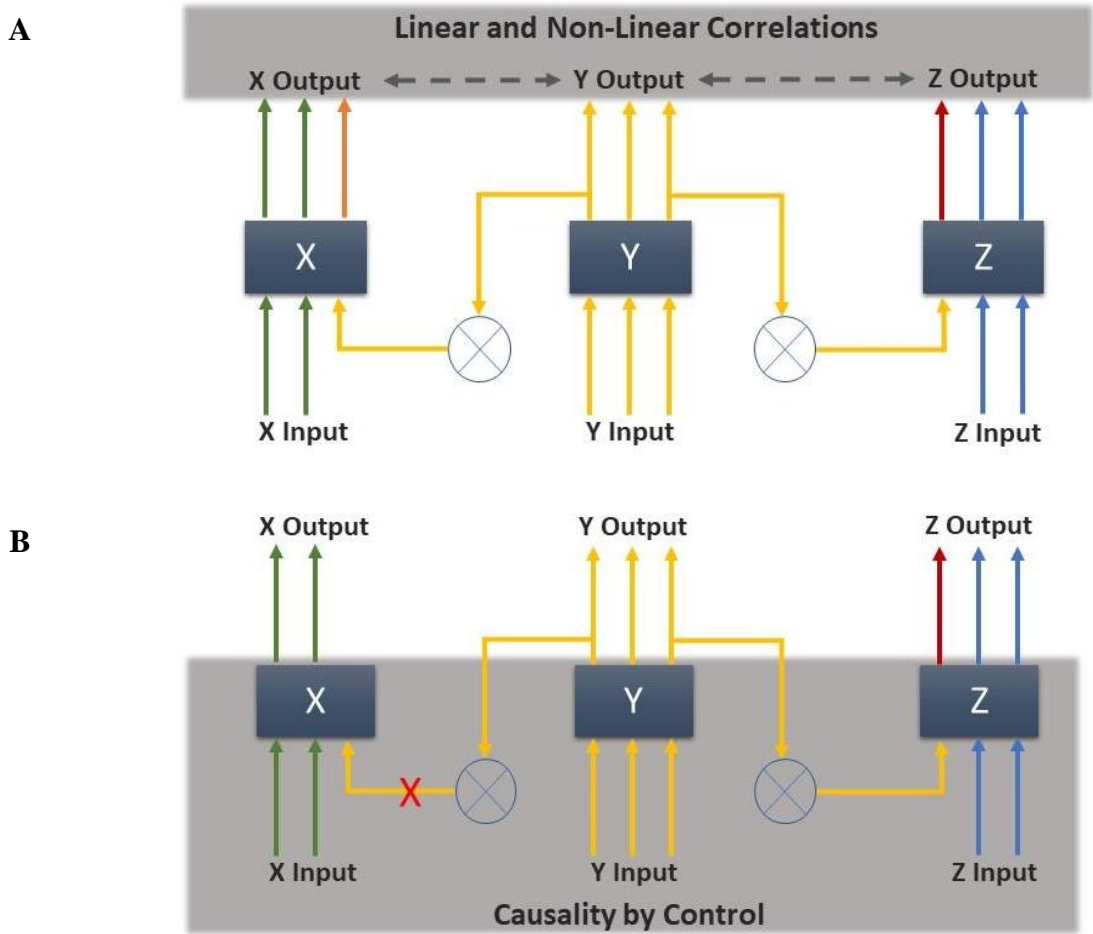

**Figure 3.1. Differences between functional and effective interaction.** (A) When there are only output signals from two or three systems it is difficult to infer causality and correlations can be more adequate to describe interaction among these systems. The measures of the output from a system in specific time windowing, for instance, only represents a statistical dependence between the variables been evaluated, performed through linear or non-linear correlation metrics. (B) When there is a mathematical model or experimental protocol that supports the manipulation of inputs of a system in function to its outputs and outputs of other systems, it is possible to analyze their effective interactions, inferring causality.

On the other hand, there are other types of connections that refer to causal influences between nodes (Friston 2011). It means that the activity of one node *X* directly influences the activity of the other connected node *Y*. Thus, node links are directionally represented with an arrow, indicating the direction of information flow from the source node to the receiver node.

They can also be bidirectional, where both nodes are reciprocally coupled (Valdes-Sosa et al. 2011).

In connectivity analysis, two main definitions of causality to approach this situation are used. The first one is based on control theory, where the input of one node *Y* is influenced by the output of another node *X*. Correlations cannot infer causality since it impossible to determine if the statistical dependence being evaluated is from one of the two nodes or from a third node, or if it occurs by chance (see Figure 3.1-A).

Any causal influence between any pair of nodes could happen not only directly between them but also through a third (or more) intermediate one. In this case, a node *X* is influenced by an intermediate node *Z* (or more) which is influenced by another node *Y* (Figure 3.1-B). By perturbing the system intentionally and observing the effects of the perturbation, as already mentioned, it is possible to evaluate possible causality effects. In this case, we can, for example, block chemically or remove a tract, subregion, or nerve that connects two nodes and see if the observed statistical dependencies between their activities are maintained or modified.

The second definition is commonly used to describe causal influences between two nodes by temporal influences of part of an output *X* onto an output *Y* (Granger 1969; Akaike 1968; Schweder 1970). This definition reads (Wiener 1956) "a signal *X* is said to be the cause of a second signal *Y* when the information about the past of *X* helps to determine the presence and/or future states of *Y* beyond and above the information from only the past of *Y*", (Valdes-Sosa et al. 2011). In this case, it is necessary to have a mathematical model that describes this temporal influence. There are some disagreements on the use of these approaches as real metrics for causality measurements because they can only provide evaluations of directed functional connectivity since they are defined in terms of time-lagged statistical dependencies (Razi and Friston 2016), figure 3.2.

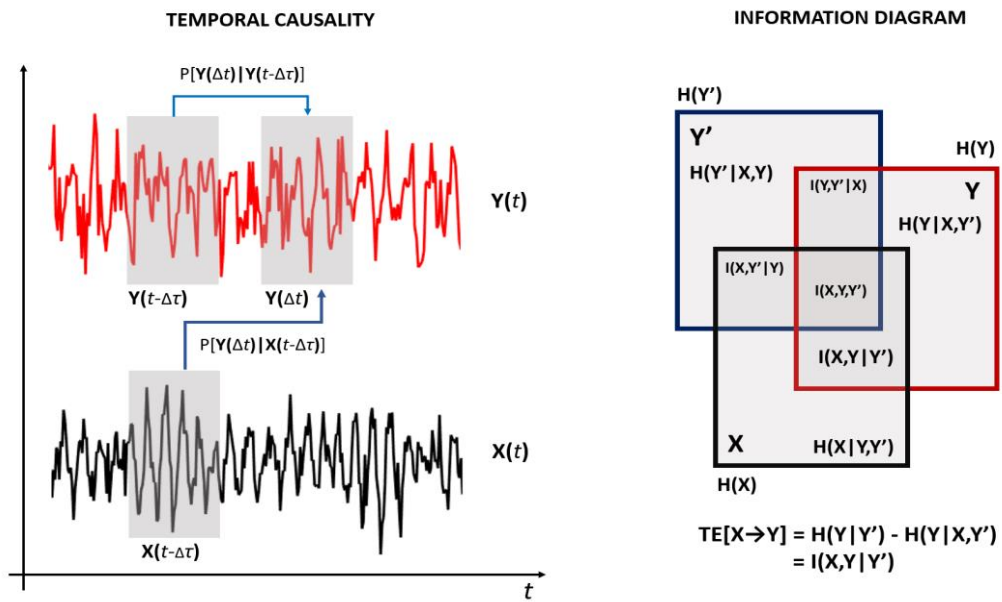

**Figure 3.2. Diagram of temporal causality and Transfer Entropy.** The definition of causality by statistical dependencies assuming temporal lags between two signals. In this definition, there is a time-windowing t-Δτ in both signals that cause the variations in y(Δt). On the right side, there is Venn diagram showing the relationship among each variable and their intersection. As case of a possible temporal causality, the Transfer Entropy metric is one of the most used in the scientific literature.

Therefore, by using these metrics, it is possible to establish a well-defined criterion to describe edges among different nodes in a network. In this way, once a correlation or a coupling metric that determine all node links is chosen, it is necessary to verify if there are spurious correlations. For this procedure, different methods can be applied as thresholds using surrogate or baseline signals, for example, in order to decide statistically significant network edges. In this way, by using the matrix representation of a graph, there are two possibilities: (*i*) the edges can be 'digitized' with those edges bigger than the threshold valued as 1s and other edges valued as 0s, or (*ii*) they can be weighted with continuous or discrete values for each edge (Papo, Zanin, and Buldú 2014).

## 3.2. Functional Connectivity

### 3.2.1. Undirected metrics

The first aspect of undirected metrics is the equilibrate flow of information between two nodes. It means that (*i*) the physical interaction between two nodes is totally symmetrical and static without flow of information from one to another node (Ahuja and Magnanti 2018); in this case, any metric provides a good measurement of the relationship between nodes; or (*ii*) there is an information flow from one node to another, but we use a symmetric statistical metric, unable to describe/detect the asymmetry between nodes. As described in Figure 2.2, this type of edge is represented by a symmetrical adjacent matrix (Wilson 1979). A commonly used metric for this type of link is a measure of correlation given by the parametric Pearson's coefficient that calculates the ratio of covariance between two variables *X* and *Y* normalized by the square root of the product of their own variance:

$$\rho = \frac{COV(X,Y)}{\sqrt{\sigma_x^2 \sigma_y^2}} \qquad (3.1)$$

*COV* (*X,Y*) represents the covariance of the variables *X* with *Y*, and $\sigma_x$ and $\sigma_y$ are their respective standard deviation. The coefficient $\rho$ ranges between -1 and 1 and depends only on the spread of *X* and *Y*, capturing only their linear correlation. In the context of connectivity, it has been applied to evaluate the degree of linear correlation of a signal's amplitude among different nodes (Lee Rodgers and Nicewander 1988). A non-parametric correlation metric is also possible, such as Kendal's and Spearman coefficient (Kelley et al. 2007; Papo, Zanin, and Buldú 2014; Meskaldji et al. 2015).

Analogously, the coherence is also a symmetric method that evaluates the spectral correlation of two signals *X*(*t*) and *Y*(*t*) in the frequency domain (Shaw 1984):

$$C_{xy}(f) = \frac{P_{xy}(f)}{\sqrt{P_{xx}(f)\, P_{yy}(f)}} \qquad (3.2)$$

Where $P_{xx}(f)$, $P_{yy}(f)$ are respectively the power spectrum of *X*(*t*) and *Y*(*t*), and $P_{xy}(f)$ is the cross-spectrum between them. The spectrum can be calculated by applying the Fourier or Wavelet transforms on the signals (Sifuzzaman, Islam, and Ali 2009). The magnitude of coherence $C_{xy}(f)$ can be normalized to values between 0 and 1, representing the intensity of the correlation power at specific frequencies (Pesaran et al. 2018). However, in order to use coherence as an interaction coefficient between two nodes, it is necessary to have an operation to summarize an index that represents the general coherence spectrum between *X* and *Y*, such as the magnitude of total coherence spectrum. It occurs because a coherence spectrum cannot be described by only one point, invalidating a network edge representation (Baccala and Sameshima 2001; Sun, Miller, and D'Esposito 2004).

The neural oscillations show specific patterns that are sometimes evidenced by the increasing or decreasing power along specific frequency bands. These fluctuations can be explored as relevant information to establish functional connectivity between nodes. In this way, spectral coherence can be used to detect aspects of phase synchronization among specific rhythms of the signals *X*(*t*) and *Y*(*t*), which may provide information on the communication dynamics among neurons (Bowyer 2016). Some researchers applied this technique to study functional networks using EEG, with electrodes placed on the scalp of a patient with a neurological disease, in order to detect its influence on the performance of different cognitive and behavioral tasks (Carmona, Suarez, and Ochoa 2017).

Spectral coherence measures an important effect of oscillations, when two or more signals have the same phase difference at a given frequency since most of the neural communication is directly related to the phase relationship of neural populations (Engel and Fries 2016).

Another way to calculate symmetrical nonlinear interaction between two nodes, *X* and *Y*, is by using the mutual information between them calculated through *I*(*X,Y*). This technique comes from the Information Theory and uses the concept of Shannon's entropy to evaluate the information shared by two or more random variables (Vergara and Estévez 2014). Considering X and Y as two random variables with specific states, $\{x_1, x_2, x_3,..., x_n\}$ and $\{y_1, y_2, y_3,..., y_n\}$ associated with their probability distributions $\{p(x_1), p(x_2), p(x_3),..., p(x_n)\}$ and $\{p(y_1), p(y_2), p(y_3),..., p(y_n)\}$, the mutual information is defined as:

$$I(X,Y) = \sum_i \sum_j p(x_i, y_j) \, log \frac{p(x_i, y_j)}{p(x_i)p(y_j)} \qquad (3.3)$$

where $p(x_i)$ and $p(y_j)$ are the probability values associated with a right state $x_i$ and $y_j$, respectively, and $p(x_i, y_j)$ is the conditional probability between these two states. Considering the definition of Shannon's entropy being the expected value of the amount of information given by some random variable, the mutual information *I*(X,Y) represents the average amount of information shared by two systems X and Y (MacKay 2003). It is important to emphasize that this metric can be used for any physical or statistical feature associated with the signals *X* and *Y*, but its representativeness is directly affected by the empirical probability distribution created to stipulate its information (Cover and Thomas 2012). One of the advantages of using *I*(X,Y) is that, since it is not a linear function, it allows generalized measurements for symmetrical statistical dependencies between two or more random variables (Bastos and Schoffelen 2016).

### 3.2.2. Directed metrics

When the interaction of any two nodes *X* and *Y* presents a privileged pathway of information flow and unsatisfied one of the two previous conditions of symmetry, the use of directional metrics is preferable to represent the network links. In general, these metrics aim to capture a statistical clue of causation or, at least, a direction of the information dynamics, considering the different degree of dependence between nodes *X* and *Y* (Bastos and Schoffelen 2016). Here, we will present three mathematical metrics that evaluate directed interaction between nodes: Kullback–Leibler (*KL*), Transfer Entropy (*TE*) and Phase Slope Index (*PSI*). However, it is important to mention that there are many other, more or less, adequate metrics according to what is intended to describe.

The Divergence of Kullback–Leibler, also known as relative entropy, is based on the concepts of information theory which can be roughly interpreted as a measure of the cost to turn a right probability distribution, P(X), into another, Q(X), under the same set of states or alphabet. Similarly, we may ask which probability distribution, P(X) or Q(X), will minimize the number of bits used to represent all the states of the random variable $X = \{x_1, x_2, x_3,..., x_n\}$? (Polani 2013; Shlens 2014). Kullback–Leibler is, therefore, an asymmetrical measurement defined as:

$$D_{KL}(P|Q) = -\sum_i p(x_i) \log \frac{p(x_i)}{q(x_i)} \tag{3.4}$$

where $p(x_i)$ and $q(x_i)$ represent the probability values from P(X) and Q(X), respectively. In a general way, the main objective to use directional metrics is to describe, in some way, statistical causations. As described previously, Transfer Entropy (TE) can be a metric that captures possible temporal causations between two signals (Figure 3.2). It appears as a new approach to contrast with the time delayed mutual information (James, Barnett, and Crutchfield 2016), measuring the information transferred between to random process, X(t) and Y(t), considering part of their past and current states (Schreiber 2006). Mathematically it is defined as:

$$TE(X \rightarrow Y) = \sum p(y_{n+1}, y_n^{(k)}, x_n^{(l)}) \log \frac{p(y_{n+1}| y_n^{(k)}, x_n^{(l)})}{p(y_{n+1}| y_n^{(k)})} \tag{3.5}$$

where $y_{n+1}$ it is a future state of Y; $y_n^{(k)}$ is an vector of k previous possible states of Y ($y_n^{(k)} = (y_1, y_2, ..., y_{n-k+1})$); $x_n^{(l)}$ is $l$ previous states of X with the minimum of $1$ and maximum of $k$ ($x_n^{(l)} = (x_1, x_2, ..., x_l)$). TE can, therefore, represent the directed information flow from X(t) to Y(t), or it can be interpreted as a degree of dependence between X and Y (Bossomaier et al. 2016).

Although Wiener had initially used the concept of causality to interpret *TE*, it has currently been claimed that *TE* is an approach to quantify the predictive information flow (Lindner et al. 2011).

Finally, an alternative technique to infer asymmetrical information flow from two signals is the phase slope index (PSI). PSI is also a nonlinear metric designed to measure the frequency-average of the slope phase of the spectral coherence (Nolte et al. 2008). As any interaction requires time and different interactions have, in general, different communication speeds between the sender and receiver, the phase difference between the sent and received messages should be assessed by the frequency. It means that PSI can detect positive or negative slopes on phase frequency-range that indicates the direction of information flow. If this relationship is negative, then the information flow occurs in the opposite direction (Cohen 2015; Maris, Fries, and van Ede 2016). The PSI is defined as:

$$\widetilde{\Psi}_{ij} = \zeta \left( \sum_{f \epsilon F} C_{xy}(f) \, C_{xy}(f + \delta f) \right) \tag{3.6}$$

where $\zeta$ represents the use of just the imaginary part of the complex number; $C_{xy}(f) = \frac{P_{xy}}{\sqrt{P_{xx}(f) \, P_{yy}(f)}}$ the spectrum coherence between $X$ and $Y$ and $\delta$ represents the frequency resolution of the spectrum. Therefore, PSI estimates the degree of coherent communication between two or more nodes. Worthy of emphasis is that if *X* has impact on *Y*, it does not imply that *Y* has no impact on *X*.

## 3.3. Effective connectivity

As already discussed, in different references, a measure of correlation does not imply a measure of causality. This relationship is at the heart of the difference between the concept of functional connectivity and that of effective connectivity.

It is not the purpose of this chapter to describe all the metrics of effective or structural connectivity, but to discuss some aspects of effective connectivity. The most common applications of effective connectivity are found in macroscopic networks through EEG, MEG, and fMRI recordings, although there is no formal restriction for using this approach (Razi and Friston 2016).

Effective connectivity is an alternative measure or extension of functional connectivity; thus, one can also minimally infer causal relationships between them (Friston 2011). The techniques used in the characterization of effective connectivity are extremely general and allow a description of connection at any scale of time and space in the neural domain.

There are currently two main mathematical approaches to describe effective connectivity between two or more nodes: The Granger Causality Modeling (GCM) and the Dynamic Causal Modeling (DCM) (Marreiros et al. 2010; Kuznetsov 2013; Friston et al. 2016).

The technique for measuring effective connectivity described by GCM was described by Granger in 1969 as a particular case of the Wiener definition given in 1956. Subsequently, other variations emerged (Akaike 1968; Schweder 1970; Valdes-Sosa et al. 2011). The causality of Granger considers only information shared by linear interactions between the signals being evaluated. Technically, the GCM uses linear multivariate autoregressive models (MVARs) from a discrete set of differential equations (Brockwell and Davis 1998; Haufe et al. 2013):

The GCM metric has been widely used in different approaches; however, caution should be taken when interpreting results because the technique requires many restrictions and conditions. The first of these is independence from random fluctuations in signal relative to past events. The second is inherent in the linearity of its expression.

Dynamic Causal Modeling (DCM) (Marreiros et al. 2010) can be understood as a more general measure than GCM because it makes less demands on the description of signal interaction models. In addition to using a continuous time formulation, it can also use a bilinear approximation (or Taylor expansion) in its interaction model (Klaas Enno Stephan and Friston 2010).

The equations described in the DCM interaction models can also be associated with a second set of time-independent equations, which map the variable X, associated with neural states, into recorded signals such as EEG, MEG or BOLD. From this projection, a generalization is made about the linearities imposed by the model, thus overcoming the restrictions found in the GCM. However, a level of caution is also important since these same projections may also insert false positives about the relationships between two nodes (Haufe et al. 2013; Klaas Enno Stephan and Friston 2010; Razi and Friston 2016).

## 4. GRAPH MEASURES

Once nodes and the connections are defined and characterized, we can use mathematical metrics to quantify network properties that define how the graph elements interact and how they are organized in time and space, forming complex mathematical structures and, finally, how these structures express storage, processing, transmission and organization of the neural information.

There are two main types of measures that characterize a graph; topological and geometric. Topological measures quantify the association between nodes irrespective of their physical location. A node can connect to another node, close to or far from it. Geometric measures, instead, evaluate how nodes are physically associated in a geographic space. Since geometric measures are usually related with the distance between connected nodes, they are frequently used in structural connectivity descriptions. For this reason, we will not discuss geometric metrics in this text, but it can find in Bullmore and Basset (2011).

### 4.1. Topological measures

Topological measure comprises a set of metrics used to quantify different network information. To choose the most informative metrics, it is necessary to consider the type of network and the type of information being studied. Essentially, topological metrics in neuronal networks measure how functional integration, segregation, efficiency, resilience, and motifs describe the information storage, flow, and assessment in the brain and how the network integrity is maintained.

Some basic network characteristics greatly affect many topological measures, like number of nodes, number of edges, clustering coefficient, path length, and degree (which is considered a fundamental measure - Figure 4.1). The degree distribution, i.e., the probability distribution of the number of edges connected to a node, can determine the complexity of arranged network architectures.

### 4.1.1. Functional segregation

In general, functional brain connections present features of complex networks with non-random connections and shared relationships (Wig, Schlaggar, and Petersen 2011). The study of brain functional segregation analyzes how networks are organized in specialized cores for information storage and processing (Figure 4.1). The best example of this organization is the brain cortex that, despite its apparent homogeneity, is composed by many distinct functional areas such as the visual and somatosensory-motor regions (Honey et al. 2007). Modules are characterized by a large number of connections between elements inside them, also called "*community*", and lower numbers among elements of other communities. Metrics able to detect these communities are known as modularity and clustering or yet module coefficients ($C$, Olaf Sporns 2013a; Rubinov and Sporns 2010). All mathematical metrics and definitions presented in this section can be also consulted in Rubinov and Sporns (2010), Bullmore and Sporns (2009), and Kaiser (2011).

In general, brain networks present higher $C$ (Ravasz & Barabási, 2003), most probably due to the network arrangement that divides and compartmentalize the information flow to optimize the mechanism of information processing. The study of how these modules are connected leads to the understanding of the integration between them. For example, how the cortical infrastructure supports a single function involving specialized areas linked by the functional integration between them. Therefore, segregation only makes sense in functional integration context and vice versa (Friston 2011; Cornelis J. Stam and Reijneveld 2007).

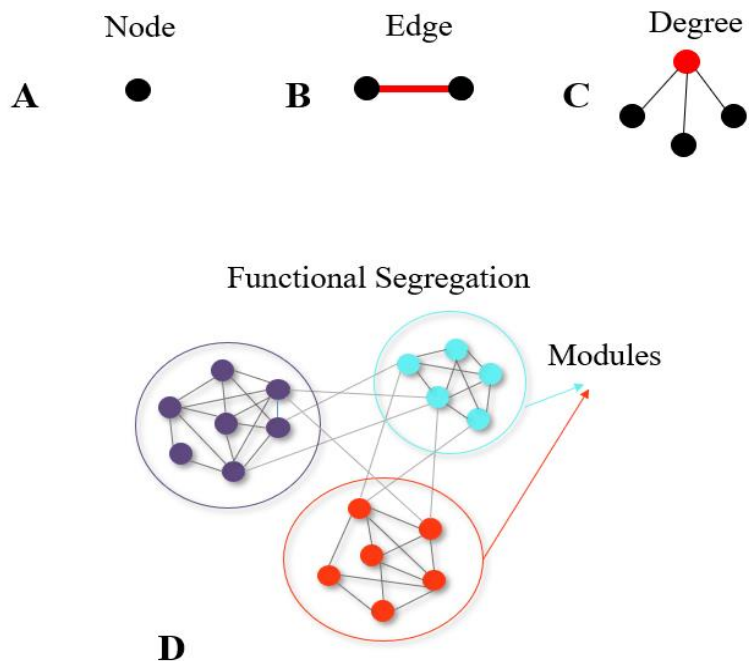

**Figure 4.1 Basic topological features and segregation measures.** (A) Node, the basic connectivity unity of a graph. (B) Edges, the connection representation between nodes (C) A node degree is the number of connections that a specific node makes with other nodes. (D) Segregation measures like clustering coefficient calculate the existence of specialized modules of information storage and processing.

### 4.1.2. Functional integration

Functional integration is associated with the network capacity to involve global interactions transcendent to the limits of modules. A cortical structure supporting or dedicated to a special function is made of many segregated information, implying that there is a coordinated activation of many neurons in different regions. Complex dynamics of cognitive or behavioral control requires efficient communication among diverse modules and a high capacity to integrate distributed information (Olaf Sporns 2002). For any connectivity analysis, measures of these attributes are associated with paths and hubs of a network.

As mentioned in Session 3, in functional connections of networks, paths are sequences of statistical dependencies that does not necessarily correspond to structural connections (Rubinov and Sporns 2010). Pathlengths indicate the efficiency between connections of different modules. A shorter path implies a stronger integration since information transmission is faster. The average shortest path length ($L$) plays an important role in the characterization of a graph and is an important measure of integration and efficiency.

Therefore, the global efficiency of a network is given by the average of the inverse shortest path length (Boccaletti et al. 2006). Based on this measure, highly disconnected networks present paths tending to infinity and, consequently, the efficiency tends to zero. In addition, there are two more network attributes that indicate the integration of information: network hubs and interconnection propensity of hubs (Olaf Sporns 2013a).

Hubs are associated with central nodes in the network and are very important due to their high connectivity density with other nodes. They can be identified by different metrics that measure degree, number of connections between specific nodes (Figure 4.1), and centrality or participation in modular connectivity. Global hubs are responsible for intermodular communication and integration, while hubs inside a module promote the cohesion of their own communities (Olaf Sporns 2013a; Kaiser 2011; Rubinov and Sporns 2010).

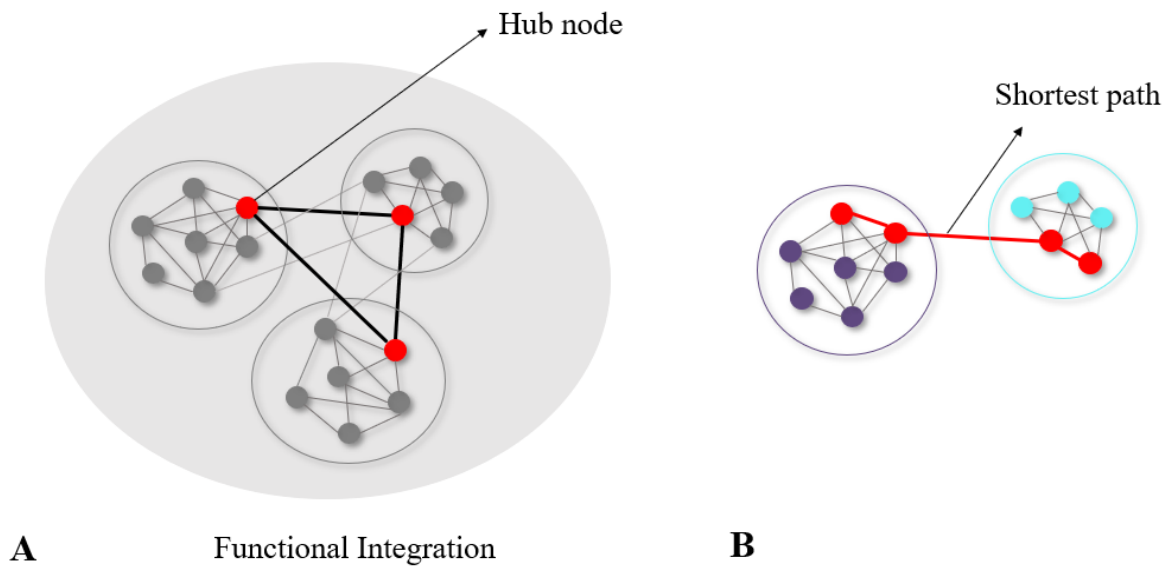

**Figure 4.2. Integration Measures.** (A) Hubs are network elements that integrate different modules and can be calculated by metrics like degree, centrality or participation in modular connectivity. (B) Paths are sequences of minimum links between distinct nodes and are associated with the network efficiency by calculating the shortest path.

### 4.1.3. Network resilience

An important network characteristic is its reliability, measured by its resilience. In brain networks, reliability is associated with the capacity of brain systems to overcome a pathological attack by a disease or an aberrant development (Danielle Smith Bassett and Bullmore 2006). Functional damages are closely related to damages in biological network structures since many neuropathologic lesions can affect functional brain activity. For instance, depending on the injured areas, some brain functions can be lost after a stroke (Khadem, Hossein-Zadeh, and Khorrami 2016). Another example is the disconnection hypothesis in schizophrenia that suggests that an impaired neuromodulation of synaptic plasticity results in an abnormal functional integration of particular neural systems (Klaas E. Stephan, Friston, and Frith 2009). The brain network resilience, therefore, is its capacity to adapt and maintain its functionality over physiological adversities, and it can be directly or indirectly characterized through topological measurements.

Network resilience can be measured, for example, by using *assortativity* or average neighbor degree (Olaf Sporns 2011; E. Bullmore and Sporns 2009a; Rubinov and Sporns 2010) before and after an insult to test the network vulnerability or the network ability to recover from that insult (Barker, Ramirez-Marquez, and Rocco 2013). Moreover, the insult can be computationally simulated by removing random or targeted nodes. Thus, the effects of the lesion may be measured and compared with the structural, functional, and effective connectivity (Rubinov and Sporns 2010).

### 4.1.4. Complex network architectures

Natural networks present architectural features that reflect their construction or development processes and function. As mentioned in section *4.1.2.*, $k_i$ is the number of vertices linked with the node *i*. Thus, the probability $p_k$ is defined as the fraction of vertices that have degree *k* in the network. In other words, $p_k$ is the probability that a node chosen at random has a degree *k* and it is given by the distribution function *P(k)* (Costa et al. 2007; Boccaletti et al. 2006; Newman 2003). The degree distribution can be presented as a histogram of the degrees of vertices and described by the function that fit the histograms as shown in Figure 4.3.

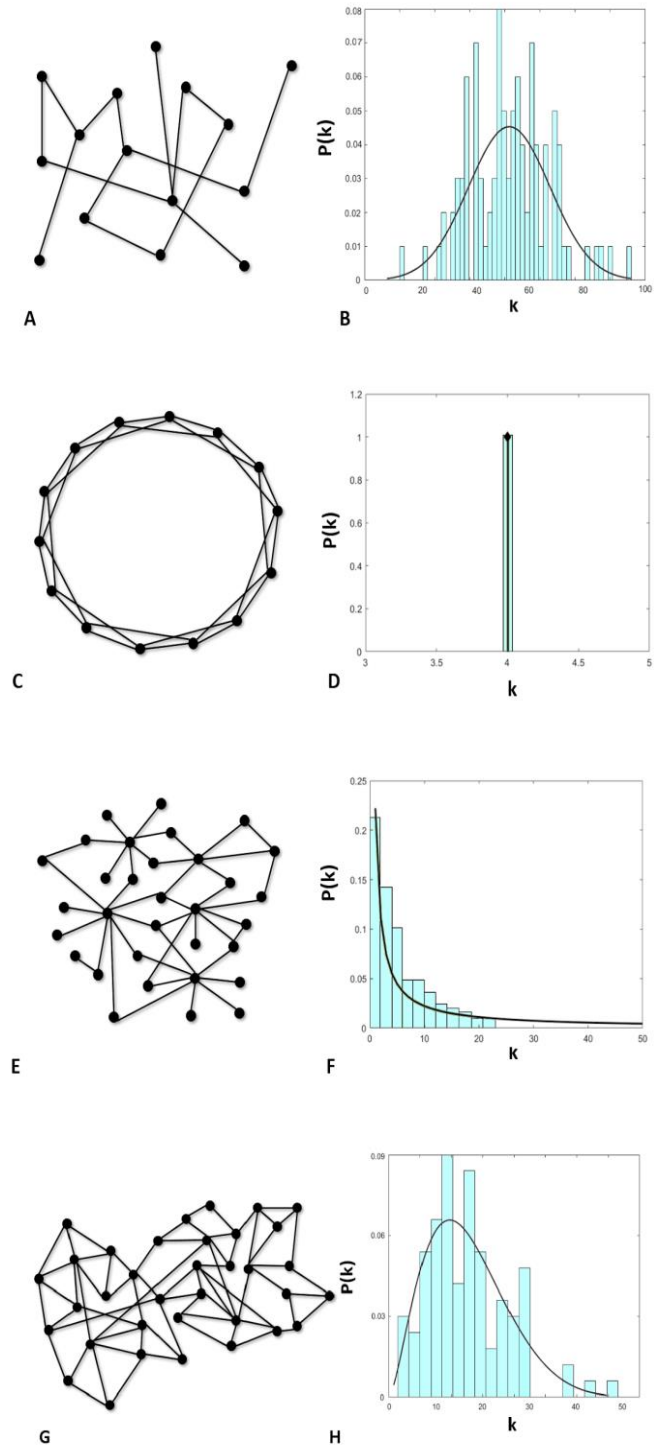

**Figure 4.3 Network architectures and examples of P(k) distributions** (A) Random network, nodes were random connected by the edges. (B) Example of a P(k) distribution of a random network with k varying from 0 to 100. (C) Lattice network, nodes were ordered connected and all of them have the same node degree. (D)Example of a P(k) distribution of the lattice network presented in item C. (E) Scale-free network, the clusters formation can be observed. (F) Example of a P(k) distribution of a scale-free network, the function of distribution follows a power law. (G) Small-world network, the formation of clusters can be visualized, and there is integration among them. (H) Example of a P(k) distribution of a small-world network with k varying from 0 to 50, the P(k) distribution may be similar to the random network distributions, what will differentiate it will be its segregating properties. All the P(k) distributions are represented by an empirical histogram and by an analytical model (Adapted from Vella et al. 2017; Costa et al. 2007).

In networks architectures, *random graphs* are associated with the disordered nature of the links between nodes. In random graphs all connections are equally probable, resulting in a Gaussian degree distribution (E. Bullmore and Sporns 2009; Boccaletti et al. 2006; Wang 2002):

$$P_k = e^{-<k>} \frac{<k>^k}{k!} \qquad (4.1)$$

where $<k>$ is the average degree of the network and $P_k$ gives the probability to randomly select a node with exactly $k$ edges. In this kind of graph, all edges are randomly designated as nodes pairs. Because of this, its $C$ and $L$ are very small and very short, respectively (Cornelis J. Stam and Reijneveld 2007).

As opposed to random graphs, *regular lattice graphs* have a very ordered pattern of connection between nodes, such as in ring or grid lattice (Figure 4.3), where the connected nodes tend to have the same neighbors but the path lengths between them vary greatly and the shortest paths are compounded by many intermediate nodes. Hence, lattice graphs have bigger $C$ and longer $L$ values (Olaf Sporns 2011). Many natural networks have an uneven distribution with a much skewed and slower decaying than a Poisson distribution. One example of this dynamics is given by a power law decaying (Strogatz 2001):

$$P_k \sim k^{-\gamma} \qquad (4.2)$$

These networks are called *scale-free* (Olaf Sporns 2011). A general characteristic of this kind of network is the existence of hubs, since some nodes are highly connected while others have few connections (Costa et al. 2007). Some examples of scale-free networks are metabolic networks (Rajula, Mauri, and Fanos 2018), gene regulatory networks (Ouma, Pogacar, and Grotewold 2018), World-Wide Web (Broder et al. 2000), etc.

Some studies have investigated a possible scale-free organization of functional connectivity in human brains, but the results of these studies have been inconclusive (He et al. 2010; Eguíluz et al. 2005). Conversely, van den Heuvel *et al.* (2008) suggested that the functional connectivity of the human brain is a combination of the scale-free and small-world organization.

*Small-world networks* combine high levels of local clustering among network nodes and short paths that globally connect all nodes of the network, promoting integration between clusters (E. Bullmore and Sporns 2009b; Danielle Smith Bassett and Bullmore 2006). As shown in Figure 4.3, both degree distributions of small-world and random networks can be fitted and modeled by a Poisson function. But the difference between them is that $C$ is much higher in a small-world network than in a random network, whereas $L$ is similar in both networks, given that they have the same size (Costa et al. 2007). Then, these criteria are used to determine a network with a small-world architecture evaluated by the expression below (Kaiser 2011):

$$S = \frac{\frac{C}{C_{rand}}}{\frac{L}{L_{rand}}} \qquad (4.3)$$

where *C* and *L* are the clustering coefficient and characteristic path length, respectively, being compared to the *C* and *L* of their corresponding random network (modeled computationally).

Watts and Strogatz (1998) demonstrated the presence of small-world topology in the nervous system of C. elegans (Hallquist and Hillary 2018; Olaf Sporns et al. 2004; Olaf Sporns and Zwi 2004). According to Basset and Bullmore (2006), there are some reasons for neural networks to be small-world, since the brain is composed by a complex network with multiple spatial and time scales. In the macro-scale network information is segregated and distributed and then is integrated to form a unique function. Similarly, small-world architecture comprises a high clustering coefficient and a short path length indicative of segregated and distributed processing and information integration, respectively. In addition, during brain development, the network is optimized to minimize costs and maximize the efficiency of information processing. These characteristics can be found in small-world networks with high global and local efficiency, which can indicate parallel information processing, low wiring costs, and sparse connectivity between nodes.

Considering all concepts together, when analyzing a functional and effective network architecture, it is important to pay attention to some critical points. For instance, depending on the feature being considered as a node and the metric to compute the communication among them, a totally different topology may arise. In this way, it is critical to be careful with the classification of specific regions of the brain, but mainly what physical/signal features, and also which metric are been considered.

A functional network described by one specific feature and specific metric can present one kind of architecture that is impossible to generalize to all functional brain networks associated with other features and metrics since the node degrees can change completely. Because of the popularity of connectomics, many studies have presented strong claims about brain networks; however, a bit of caution and conservatism is needed since it is very difficult to reduce all aspects of the brain to simple features and simple interactions.

### 4.1.5. Network motifs

Significant and recurrent patterns of node interconnections are known as network motifs. Usually, the connection patterns of a network are compared with a random network to find patterns that appear in numbers significantly higher than those in a randomized network (Milo et al. 2002). In this way, network motifs are well-defined connectivity blocks that appear in a right network with equal or greater probability when compared with a random network simultaneously lower than a cutoff value. It is important to mention that there are patterns without any statistical significance that are still important for the network (Milo et al. 2002).

As shown in Figure 4.5, the functional network topology can exhibit, for instance, triangles and feedback loops or biparallel blocks that represent specific mechanisms of the network such as information protection, processing, and storage. The network motifs can also be measured by its frequency of occurrence, normalized as the motif z-score (Rubinov and Sporns 2010).

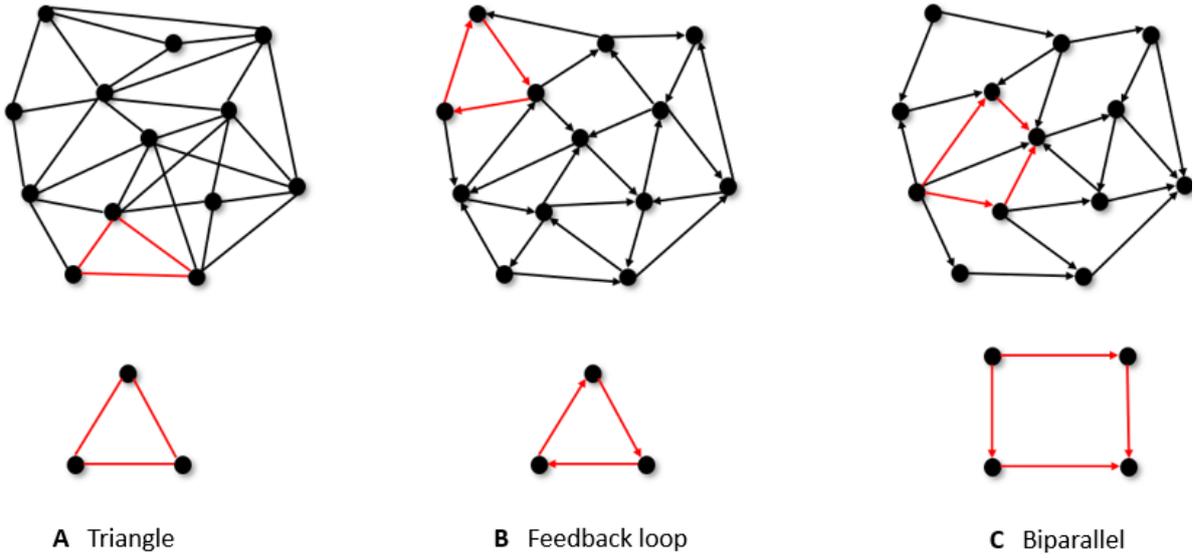

**Figure 4.5 Motifs examples.** Three recurring connection patterns are show: (A) three undirected edges form a triangle linking three nodes, (B) three directional edges link three nodes forming a feedback loop and (C) four edges form two parallel ways that leave and arrive at the same node.

| Measure | Description | Mathematical definition |
|---|---|---|
| *Node degree* | Measures the number of connections of node *i*. It is a basic measure used by many others measures. | $$k_i = \sum_{i,j \in N} a_{ij}$$ $a_{ij}$ is the connection status between the node i and the node j (when i and j are neighbors). When the link exists the $a_{ij}$ value is 1 otherwise is 0. |
| *Clustering coefficient* | Measures the degree that the graph nodes tend to cluster together. | $$C_i = \frac{\Gamma_i}{k_i(k_i - 1)}$$ $\Gamma_i$ is edges between neighbors |
| *Global clustering coefficient* | Measures the clustering coefficient of the entire network. | $$C = \frac{1}{n}\sum_{i \in N} C_i$$ *n* is the number of nodes in the network and $C_i$ is the Cluster coefficient. |
| *Modularity* | Measures the strength that a network is divided into modules. | $$Q = \frac{1}{l}\sum_{i,j \in N}\left[a_{ij} - \frac{k_{iin} - k_{jout}}{T}\right]\delta_{c_ic_j}$$ *l* is the total number of edges, $a_{ij}$ is the element of the adjacency matrix, $k_{iin}$ is the degree of node i, $k_{jout}$ is the degree of node j, $\delta_{c_ic_j}$ is the Kronecker delta (1 if nodes i and j are in the same module and zero, otherwise). |
| *Shortest path length* | Measure of the shortest path length between two nodes. | $$d_{ij} = \sum_{a_{kj} \in g(k \leftrightarrow j)} a_{kj}$$ $a_{kj}$ is the the connection status between nodes in the shortest path (geodesic distance) between nodes k and j ($g(k \leftrightarrow j)$, $a_{kj}$= 1 if there is connection and 0 otherwise). |
| *Average shortest path length* | Measure the average shortest path length between all nodes in a network. | $$L = \frac{1}{n(n-1)}\sum_{i,j \in N} d_{ij}$$ *n* is the total number of nodes in the network. |
| *Global efficiency* | Measure of how efficiently the network globally exchanges information | $$E = \frac{1}{n(n-1)}\sum_{i,j \in N}\frac{1}{d_{ij}}$$ $d_{ij}$ is the shortest path length and *n* is the total number of nodes in the network |
| *Closeness centrality* | Measure of centrality that indicates the average length of the shortest path between a node *i* and all other nodes in the network. | $$L_i^{-1} = \frac{n-1}{\sum_{j \in N, j \neq 1} d_{ij}}$$ *n* is the total number of nodes in the network. (Normalized form) |
| *Betweenness centrality* | Measure of centrality that indicates how many times a node acts as a bridge along the shortest path between two other nodes. | $b_i = \dfrac{1}{(n-1)(n-2)} \sum_{k,j \in N} \dfrac{\rho_{hj}(i)}{\rho_{hj}}$<br><br>$\rho_{hj}$ is the number of shortest paths between h and j, and $\rho_{hj}(i)$ is the number of shortest paths between h and j that pass through i. |

From the point of view of brain complexity, where structural connectivity generates different functional states with different features, Sporns & Kotter (2004) hypothesized that brain networks have a large number of motifs in functional connectivity to maximize their number and diversity of cognitive states, keeping the same number of structural motifs.

## 5. NETWORK DYNAMICS AND MULTILAYER NETWORKS

As previously described, the structural brain connectome can be modeled as networks at different spatial scales (Olaf Sporns, Tononi, and Kötter 2005; Olaf Sporns 2013b; Horn et al. 2014). However, how the structural and functional connectivity interplay with and within other neural networks in space and time remains unclear. The investigation of this question embraces multiple spatiotemporal scales and demand several modalities of experimental techniques for data recording. The multilayer network approach allows for the merging of datasets in a consistent way and bring to light the hidden features of the complex organization of the brain networks.

### 5.1. From Static to Dynamic Networks

As described in section 2, a network is defined as a graph (G), an abstract representation corresponding to a network arrange. To capture further information in the graph, such as temporal information, new edges need to be included along additional non-nodal dimensions. The extended mathematical definition of a graph (see section 2.1) is:

$$G = \{V, E, D\} \qquad (5.1)$$

where *D* is a set of additional non-nodal dimensions (Thompson, Brantefors, and Fransson 2017). The Equation 5.1 is sometimes referred in mathematics as a multigraph network and, in network theory, as a multilayer or multiplex network (Kivelä et al. 2014).

A graph is said to be a dynamical network (or a temporal network) when *D* contains an ordered set of temporal indices representing time. In this way, *D* could be a set containing discrete temporal indices in seconds, minutes, hours, days or even years: $t = \{1, 2, \ldots, T\}$. In this case, a temporal network can be expressed by a group of static graphs $G = G^t$ of size *NxN* nodes, corresponding to a series of "photographs" of the network at each time *t* (Figure 5.1, Dai et al. 2016; Thompson, Brantefors, and Fransson 2017).

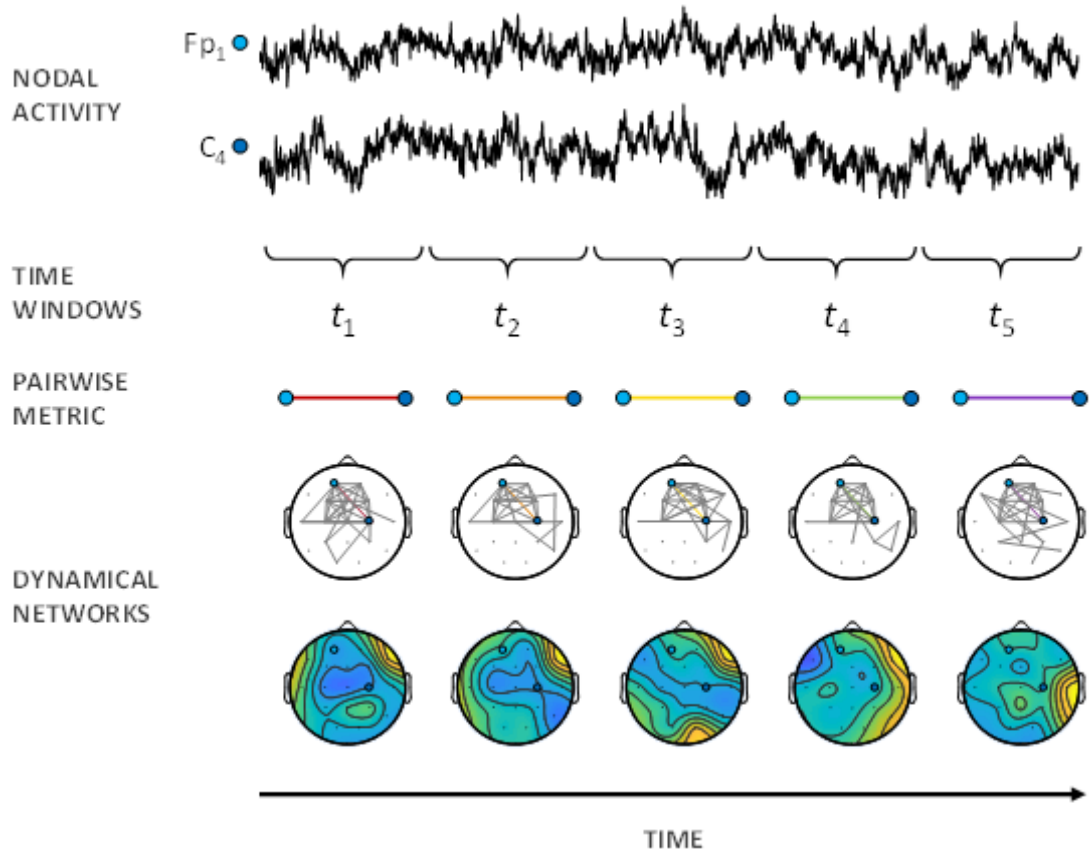

**Figure 5.1. Scheme of dynamical networks in the brain.** In a dynamic functional connectivity analysis, the nodal time series are time-windowed and the relationship between pairs of nodes is given by connectivity metrics (see the measures described in sections 3.2 and 3.3). The static layers for each time-window ($G^t$) are concatenate to a $NxNxT$ array representing the changes in functional connectivity between nodes as a function of time.

The dynamical networks theory can be applied to investigate the oscillatory behavior of the resting state networks, since it allows us to assess the intrinsic dynamics of the network nodes and the couplings between all nodes of a network across time, revealing the functional network dynamics (Deco, Jirsa, and McIntosh 2011).

The alterations in brain connectivity over time can be assessed by multiple approaches (Calhoun et al. 2014), such as: (*i*) the dynamic functional network connectivity: characterization of temporal coupling changes between fixed spatial networks (Doron, Bassett, and Gazzaniga 2012); (*ii*) the time-varying spatial connectivity: evaluation of changes in spatial patterns of correlated networks over time (Ma et al. 2014); and (*iii*) the time-varying graph metrics: quantification of graph measures which can reconfigure over time (Calhoun et al. 2014). Thompson at al. (2017) presented the principal measures from dynamical network theory (briefly described in Table 5.1). Most of the mathematical metrics are applied to binary, non-directed graphs, although several of them can be fitted for non-binary, directed, and continuous time graphs.

**Table 5.1 Principal dynamical network metrics.**

| Measure | Description | Mathematical definition |
|---|---|---|
| *Local level measures* | | |

| | Influence of a node $i$ in the dynamical network, calculated by the sum of the edges for the node and the number of the edges across time. | $D_i^T = \sum_{j=1}^{N} \sum_{t=1}^{T} G_{i,j}^t$ |
|---|---|---|
| Temporal centrality ($D^T$) | | $T$: number of time points<br>$N$: number of nodes<br>$G_{i,j}^t$: time-graphlet |
| Temporal closeness centrality ($C^T$) | Time between connections, given by the inverse sum of the shortest paths across all time points, between nodes $i$ and $j$. | $C_{i,t}^T = \frac{1}{N-1} \sum_{j=1}^{N} \frac{1}{d_{i,j}^t}$<br>$d_{i,j}^t$: average shortest path |
| Burstiness ($B_{ij}$) | Measure of distribution of subsequent connections per edge. $B > 0$ indicates that the temporal connectivity is bursty | $B_{ij} = \frac{\sigma(\tau_{ij}) - \mu(\tau_{ij})}{\sigma(\tau_{ij}) + \mu(\tau_{ij})}$<br>$\tau_{ij}$: distribution of intercontact times between nodes $i$ and $j$ through time<br>$\sigma$: standard deviation, $\mu$: mean |

*Global level measures*

| | | |
|---|---|---|
| Fluctuability ($F$) | Ratio of number of edges present in $G$ over the all edges of $G^t$. Quantify the temporal variability of connectivity. $F$ is 1 when every edge is unique and occurs only once in time. | $F = \frac{\sum_i \sum_j U(G_{I,j})}{\sum_i \sum_j \sum_t G_{I,j}^t}$<br>$U(G_{i,j}) = 1, if \sum_t^T G_{i,j}^t > 0$<br>$0, if \sum_t^T G_{i,j}^t = 0$ |
| Volatility ($V$) | Rate of consecutive change of graphlets over time | $V = \frac{1}{T-1} \sum_{t=1}^{T-1} D(G^t, G^{t+1})$<br>$D$: Hamming distance that quantifies the difference between a graphlet at $t$ and the graphlet at $t+1$ |
| Reachability latency ($R_r$) | Quantifies the average time it takes for a dynamical network to reach an *a priori* defined reachability ratio $r$ | $R_r = \frac{1}{TN} \sum_t \sum_i d_i^t$<br>$d_i^t$: ordered vector of length $N$ of the shortest temporal paths for node $i$ at time point $t$.<br>$k$ : $[rN]$th element of $d_i^t$ |
| Temporal efficiency ($E$) | Inverse average shortest temporal path. This measure is calculated at each time point through the inverse of the average shortest path length for all nodes to obtain an estimate of global temporal efficiency | $E = \frac{1}{T(N^2 - N)} \sum_{I,j,t} \frac{1}{d_{I,j}^t},$<br>$i \neq j$ |

The burstiness coefficient ($B_{ij}$) in Table 5.1 calculates the number of bursts per edge but can also be applied to a given node by the summation of the burstiness coefficient of all edges associated with its node. Similarly, the definitions of fluctuability, volatility, and temporal efficiency can be extended to a nodal level.

The set of metrics described in Table 5.1 summarize the connectivity information over short- and long-term time-scales, allowing to identify groups of edges that have similar temporal evolution or investigate how different tasks evoke different network configurations (Cole et al. 2013; Davison et al. 2015). However, it is necessary to evaluate which dynamic network metrics are more appropriate for each research problem. Although dynamical network theory allows access to several metrics, it is not advisable to apply all the available measures to a given dataset. A hypothesis about a possible network state should first be considered and after a measure that will help to quantify this network configuration and why it is considered. Besides, for the interpretation of a specific measure, an account of the data temporal resolution and the level of network organization under analysis must be taken, globally or locally, (Thompson, Brantefors, and Fransson 2017).

Investigations are still needed to validate existing models, build improved models, and develop high-level summary metrics. In addition, there are still several aspects of time-varying brain connectivity that need to be studied, which include: (*i*) mathematical or physical models that can capture both spatial and temporal couplings; (*ii*) approaches to capture both static and dynamic connectivity; and (*iii*) application of the existing tools to large datasets to identify predictor parameters.

## 5.2. Multilayer networks

The multilayer network formalism allows us also to include many other dimensions of information, encoding its different network layers (Betzel and Bassett 2017; De Domenico, Sasai, and Arenas 2016; Vaiana and Muldoon 2018), including: (*i*) the activity across multiple spatial scales (Yakushev et al. 2018); (*ii*) the activity in different frequency-bands (Yu et al. 2014); (*iii*) the multi-modal networks connectivity (Garcés et al. 2016); and (*iv*) the relationship of structural and functional/effective networks (Battiston et al. 2017). Therefore, a multilayer network can be described as a network of networks (Agostino and Scala 2014), or a network that contains different layers, in which the edges in a given layer represent a different type of relationship in another layer (Danielle S. Bassett, Khambhati, and Grafton 2016), Figure 5.2.

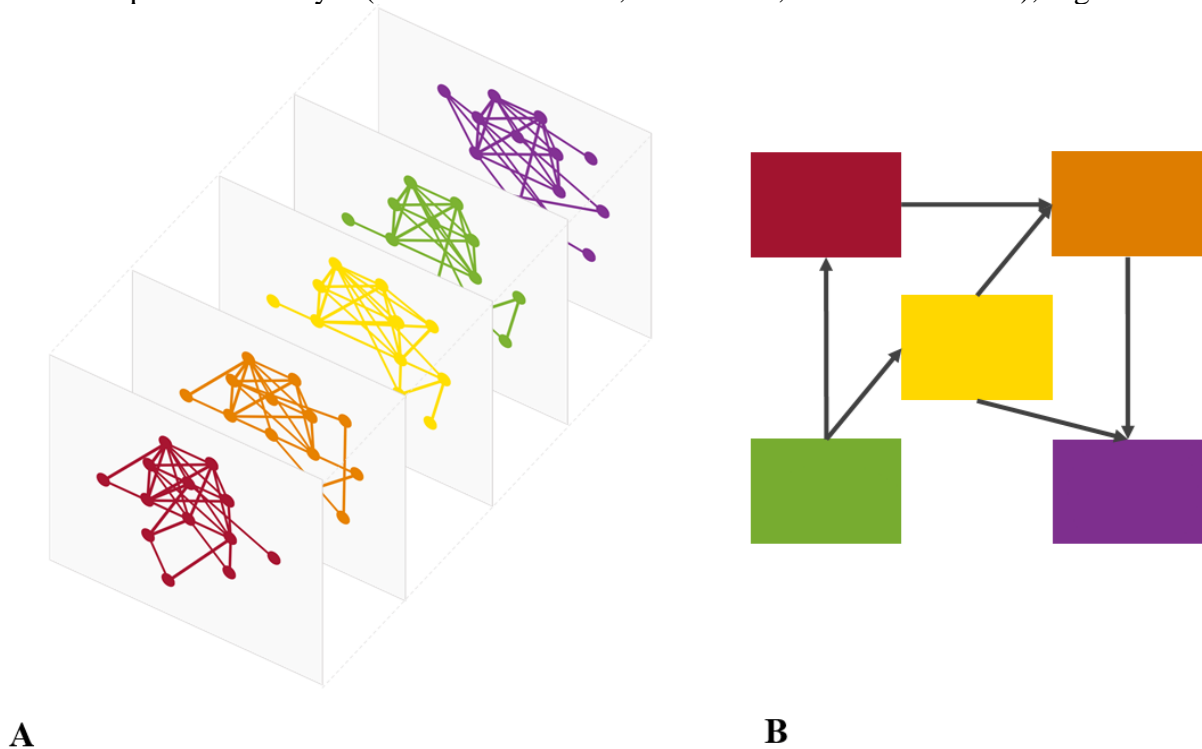

A                      B

**Figure 5.2 Schematic representation of multilayer networks.** The multilayer network framework allows to represent systems that consist of networks at multiple levels or with multiple types of edges (e.g., the functional brain activity in different spatial scales). In (A), each layer of a multilayer network corresponds to a different type of interaction between nodes and is represented by a different adjacency matrix. As shown in (B), the layers can also be interconnected. In the multilayer network approach, the interlayer edges can embrace pairwise connections between all possible combinations of nodes and layers and it is possible generalize this framework to consider hyperedges that connect more than two nodes (for details, see Kivelä et al. 2014).

The traditional network models have provided key insights into the structure and function of the brain through the assessment of descriptive and inferential network measures (Olaf Sporns et al. 2004; E. Bullmore and Sporns 2009). However, single networks provide a limited representation of the brain structure by excluding or aggregating the multiple connection types between its components (Newman 2003; Kivelä et al. 2014). The multilayered network approach for modelling brain organization allows the incorporation of multiple structural relationships, known as multiplexity (De Domenico et al. 2014; Lee, Min, and Goh 2015), that go beyond the statistical dependencies or correlations between network elements (Kivelä et al. 2014; De Domenico, Sasai, and Arenas 2016). A fundamental aspect of describing multilayered networks is defining and quantifying the interconnectivity between different categories of connections. This amounts to switching between layers in a multilayered system, and the associated inter-layered connections in a network are responsible for the emergence of new phenomena in multilayered networks.

The structure of a multilayer network can be represented by a supra-adjacency matrix (Kivelä et al. 2014), as shown in Figure 5.3. This approach allows the application of numerous tools and methods that have been developed for matrices in the investigation of multilayer networks. Additionally, the supra-adjacency matrix representation is useful to describe walks on multilayer networks and provide a way to depict multilayer networks that are not node-aligned without added empty nodes (Kivelä et al. 2014). However, to construct the supra-adjacency matrix, we must flatten the multilayer network and some of the information can be lost. To overcome this issue Kivelä et al. (2014) suggested that the edge set of the multilayer network should be grouped with the intralayered edges. The interlayered and coupling edges construct the respective supra-adjacency matrix.

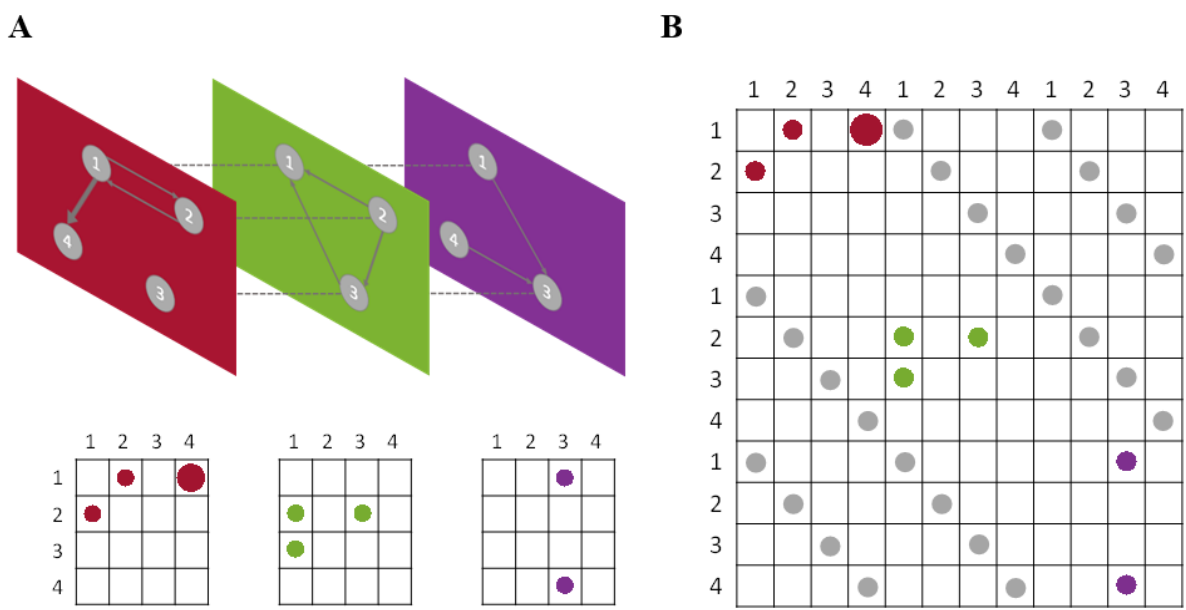

**Figure 5.3. Representation of a supra-adjacency matrix.** (A) A multilayer network constitutes of three different layers, each one represented by a (possible directed and/or weighted) adjacency matrix. (B) The supra-adjacency matrix of multilayer network shown in (A), representing intra- and inter-layer connectivity.

### 5.3. Edges between networks

In a multilayer network, establishing how each subnet communicates and connects is a job that depends on experimental factors or mathematical models that describe how each scale transmits information to another. In situations where information levels are the same between different layers of networks, describing this link is reduced to describing the communication between different modules in a complex network. However, adequately describing this link may not be such a simple task when there are different layers processing different information through nodes with different modalities.

Certainly, experimental criteria and protocols that guide the type and intensity of these communications offer great help in describing these connections. However, depending on the types of networks being connected, an experimental protocol may need levels of control that are difficult to achieve, especially when evaluating functional and effective networks in human brains. Today, this topic is one of the great frontiers in network theories due to the large number of variables and parameters to be considered as information in the description of isolated and, mainly, connected networks.

Thus, for a better description of connectivity patterns and brain activity, there is still a long way to go in order to better understand all the relationships and effects of each characteristic and each metric in different network topological measures.

## 6. CLINICAL APPLICATIONS

Technological developments of non-invasive neuroimaging techniques associated with powerful computational processing and big data have made possible the study of brain function in normal and diseased conditions. Many multinational groups have been formed with the aim of combining technologies to study the brain, such as the Human Brain Project (HBP - http://www.humanbrainproject.eu/en/), Brain Initiative (https://www.braininitiative.nih.gov/), Human Connectome (HCP - https://www.neuroscienceblueprint.nih.gov/connectome/), Virtual Brain (VB – http://www.thevirtualbrain.org/tvb/zwei), Brain Minds (http://brainminds.jp/en/central/mission), Blue Brain Project (BBP - https://bluebrain.epfl.ch/), China Brain Project (https://www.ncbi.nlm.nih.gov/pubmed/27809999), as well as many other research groups. These initiatives promote discoveries in neurological and neuropsychiatric diseases that can facilitate our understanding of their behavior and consequently their treatment.

Network theory has made it possible to understand how brain networks develop and how structures, from genes to brain areas, interact to form architectures that show universal features like hierarchical modularity and small-world organization, and how it is associated with functional cognitive and intelligence. Understanding the healthy dynamics of the brain can lead to an understanding of what has been damaged in neurological and neuropsychiatric diseases. Hence, network theory has been explored in clinical neurophysiology and has impacted clinical concepts. A good example of this is the idea that Alzheimer's disease and epilepsy can be explained in terms of "hub failure" (C. J. Stam and van Straaten 2012).

Since, in principle, a disease alters the regular functioning of the brain, we can assume that functional connectivity will also be altered. For this reason, many studies were designed to understand functional and effective connectivity under normal and diseased conditions.

Huang et al. (2018), using resting-state fMRI (RS-fMRI), studied the functional connectivity pre-to-post-operative after total knee arthroplasty with general anesthesia. They observed that 48 hours after surgery at least one fourth of the sample showed significant functional network decline, indicating that the integrity of the brain is disturbed after general anesthesia. Supekar et al. (2008) compared network topological properties of children and young-adults and concluded that apart from the fact that both networks have small-world organization, they differ significantly in hierarchical organization and interregional connectivity, suggesting the existence of key principles underlying functional brain maturation.

A very studied neuropsychiatric disease using network theory is schizophrenia. Van den Heuvel et al. (2013) examined a topology structure of *rich club* in patients with schizophrenia and its role in global functional brain dynamics. They noticed a reduction of *rich club* connections in patients and associated this reduction with lower levels of global communicative capacity. Lynall et al. (2010) measured aspects of functional network topology using RS-fMRI time series of schizophrenics. They used interregional correlation matrices to construct weight graphs and observed that the schizophrenic group showed weaker integration and more diverse connections in functional connectivity. Ganella et al. (2018) used RS-fMRI in treatment resistant schizophrenics (TRS) and unaffected first-degree family members (UFM) to study risk or resilience endophenotypes in schizophrenia associated with functional brain connectivity. They infer that both the TRS and UFM groups had functional connectivity deficits representing a risk endophenotype. Nevertheless, the UFM functional connectivity is more topologically resilient than that in TRS, which may explain the absence of schizophrenia despite familial liability.

Network theory is also widely used in the study of epilepsy. Zhang et al. (2011) hypothesized a decoupling of structural and functional connectivity in epilepsy. Using fMRI images in idiopathic generalized epileptics, they corroborated their hypothesis and suggested that this decoupling can be used as a biomarker of subtle brain abnormalities in epilepsy. Hogan (2018) studied the effects of local insults on brain development. They found indications of correlation between severity of topological network reconfiguration and time of insult during corticogenesis. Tecchio et al. (2018) also studied patients with drug-resistant epilepsy (DRE), investigating changes in functional connectivity caused by cathodal transcranial direct current stimulation (ctDCS) using eLORETA analysis in EEG data. They verified a correlation between epileptic seizure reduction and increase of functional connectivity in the epileptic focus after ctDCS in DRE patients, which can contribute to understanding the underlying mechanisms of ctDCS treatment.

Lately, Alzheimer's disease (AD) has been studied with the aid of network theory. Supekar et al. (2008) studied if the small-world brain properties are lost in AD. They noticed that the clustering coefficient distinguished health subjects into patients with sensitivity of 72% and specificity of 78%, indicating that it may be a potential biomarker of AD. C.J. Stam et al. (2007) also investigated the functional brain networks disruption in AD patients. They used beta band–filtered EEG and observed that AD patients presented a characteristic path length longer than that of healthy subjects, while clustering coefficient did not present significant changes. These findings suggest a less optimal organization and a loss of complexity in AD. Kabbara et al. (2018) reported that AD patients showed less functional integration and more functional segregation compared with healthy subjects. They also found an association between cognitive scores of AD patients and their alterations in functional brain networks.

These applications exemplify the breadth, robustness, diversity, and effectiveness that an analysis of brain dynamics based on a network approach can provide. There is still a great road to drive, and it is not yet known if graph-based network analysis will be only a new technique for quantifying patterns or whether it will be a true theory with a new perspective on how the brain represents information through its biological structures and how it processes information in time and space, integrating different modalities and providing cognitive emergent states.

## 7. REFERENCES

Agostino, Gregorio D, and Antonio Scala. 2014. *Networks of Networks: The Last Frontier of Complexity*. Networks of Networks: The Last Frontier of Complexity. Vol. 97. https://doi.org/10.1007/978-3-319-03518-5.

Ahuja, R. K., and T. L. Magnanti. 2018. "Network Flows: Theory, Algorithms, and Applications." In *Network Flows*, edited by Pearson Education.

Akaike, Hirotugu. 1968. "On the Use of a Linear Model for the Identification of Feedback Systems." *Annals of the Institute of Statistical Mathematics* 20 (1): 425–39. https://doi.org/10.1007/BF02911655.

Allen, E. A., E. Damaraju, T. Eichele, L. Wu, and V. D. Calhoun. 2018. "EEG Signatures of Dynamic Functional Network Connectivity States." *Brain Topography* 31 (1): 101–16. https://doi.org/10.1007/s10548-017-0546-2.

Altman, Naomi, and Martin Krzywinski. 2015. "Points of Significance: Association, Correlation and Causation." *Nature Methods* 12 (10): 899–900. https://doi.org/10.1038/nmeth.3587.

Amaral, L. A. N., and J. M. Ottino. 2004. "Complex Networks." *The European Physical Journal B - Condensed Matter* 38 (2): 147–62. https://doi.org/10.1140/epjb/e2004-00110-5.

Avena-Koenigsberger, Andrea, Bratislav Misic, and Olaf Sporns. 2017. "Communication Dynamics in Complex Brain Networks." *Nature Reviews Neuroscience* 19 (1): 17–33. https://doi.org/10.1038/nrn.2017.149.

Babiloni, Claudio, Vittorio Pizzella, Cosimo Del Gratta, Antonio Ferretti, and Gian Luca Romani. 2009. *Chapter 5 Fundamentals of Electroencefalography, Magnetoencefalography, and Functional Magnetic Resonance Imaging. International Review of Neurobiology*. 1st ed. Vol. 86. Elsevier Inc. https://doi.org/10.1016/S0074-7742(09)86005-4.

Baccala, L., and K. Sameshima. 2001. "Partial Directed Coherence: A New Concept in Neural Structure Determination." *Biol. Cybern.* 84 (1): 463–474. https://doi.org/10.1007/PL00007990.

Baker, Joseph M, Jennifer L Bruno, Andrew Gundran, S M Hadi Hosseini, and L Reiss. 2018. "FNIRS Measurement of Cortical Activation and Functional Connectivity during a Visuospatial Working Memory Task," 1–22. https://doi.org/10.1371/journal.pone.0201486.

Baliki, M. N., P. Y. Geha, A. V. Apkarian, and D. R. Chialvo. 2008. "Beyond Feeling: Chronic Pain Hurts the Brain, Disrupting the Default-Mode Network Dynamics." *Journal of Neuroscience* 28 (6): 1398–1403. https://doi.org/10.1523/JNEUROSCI.4123-07.2008.

Barch, Deanna M., Gregory C. Burgess, Michael P. Harms, Steven E. Petersen, Bradley L. Schlaggar, Maurizio Corbetta, Matthew F. Glasser, et al. 2013. "Function in the Human Connectome: Task-FMRI and Individual Differences in Behavior." *NeuroImage* 80: 169–89. https://doi.org/10.1016/j.neuroimage.2013.05.033.

Barker, Kash, Jose Emmanuel Ramirez-Marquez, and Claudio M. Rocco. 2013. "Resilience-Based Network Component Importance Measures." *Reliability Engineering and System Safety* 117: 89–97. https://doi.org/10.1016/j.ress.2013.03.012.

Bassett, Danielle S., Ankit N. Khambhati, and Scott T. Grafton. 2016. "Emerging Frontiers of Neuroengineering: A Network Science of Brain Connectivity," no. March: 327–52.

https://doi.org/10.1146/annurev-bioeng-071516-044511.

Bassett, Danielle S., and Olaf Sporns. 2017. "Network Neuroscience." *Nature Neuroscience* 20 (3): 353–64. https://doi.org/10.1038/nn.4502.

Bassett, Danielle Smith, and Ed Bullmore. 2006. "Small-World Brain Networks." *Neuroscientist* 12 (6): 512–23. https://doi.org/10.1177/1073858406293182.

Bastos, André M., and Jan-Mathijs Schoffelen. 2016. "A Tutorial Review of Functional Connectivity Analysis Methods and Their Interpretational Pitfalls." *Frontiers in Systems Neuroscience* 9 (January): 1–23. https://doi.org/10.3389/fnsys.2015.00175.

Battiston, Federico, Vincenzo Nicosia, Mario Chavez, and Vito Latora. 2017. "Multilayer Motif Analysis of Brain Networks." *Chaos* 27 (4). https://doi.org/10.1063/1.4979282.

Betzel, Richard F., and Danielle S. Bassett. 2017. "Multi-Scale Brain Networks." *NeuroImage* 160 (November): 73–83. https://doi.org/10.1016/j.neuroimage.2016.11.006.

Boccaletti, Stefano, V. Latora, Y. Moreno, M. Chavez, and D. U. Hwang. 2006. "Complex Networks: Structure and Dynamics." *Physics Reports* 424 (4–5): 175–308. https://doi.org/10.1016/j.physrep.2005.10.009.

Bossomaier, Terry, Lionel Barnett, Michael Harré, and Joseph T Lizier. 2016. *An Introduction to Transfer Entropy*. Springer.

Bowyer, Susan M. 2016. "Coherence a Measure of the Brain Networks: Past and Present." *Neuropsychiatric Electrophysiology* 2 (1): 1. https://doi.org/10.1186/s40810-015-0015-7.

Bradley, Allison, Jun Yao, Jules Dewald, and Claus Peter Richter. 2016. "Evaluation of Electroencephalography Source Localization Algorithms with Multiple Cortical Sources." *PloS One* 11 (1): e0147266. https://doi.org/10.1371/journal.pone.0147266.

Brockwell, P.J., and R.A. Davis. 1998. *Time Series: Theory and Methods.* Edited by Springer. *Journal of the American Statistical Association*. Vol. 92. New York. https://doi.org/10.2307/2965440.

Broder, Andrei, Ravi Kumar, Farzin Maghoul, Prabhakar Raghavan, Sridhar Rajagopalan, Raymie Stata, Andrew Tomkins, and Janet Wiener. 2000. "Graph Structure in the Web." *Computer Networks* 33 (1): 309–20. https://doi.org/10.1016/S1389-1286(00)00083-9.

Bullmore, Ed, and Olaf Sporns. 2009a. "Complex Brain Networks: Graph Theoretical Analysis of Structural and Functional Systems." Nature Reviews Neuroscience 10 (3): 186–98. https://doi.org/10.1038/nrn2575.

Bullmore, Edward T., and Danielle S. Bassett. 2011. "Brain Graphs: Graphical Models of the Human Brain Connectome." *Ssrn*. https://doi.org/10.1146/annurev-clinpsy-040510-143934.

Buzsáki, György, Costas A. Anastassiou, and Christof Koch. 2012. "The Origin of Extracellular Fields and Currents-EEG, ECoG, LFP and Spikes." *Nature Reviews Neuroscience* 13 (6): 407–20. https://doi.org/10.1038/nrn3241.

Calhoun, Vince D., Robyn Miller, Godfrey Pearlson, and Tulay Adali. 2014. "The Chronnectome: Time-Varying Connectivity Networks as the Next Frontier in FMRI Data Discovery." *Neuron* 84 (2): 262–74. https://doi.org/10.1016/j.neuron.2014.10.015.

Canolty, Ryan T., and Robert T. Knight. 2010. "The Functional Role of Cross-Frequency Coupling." *Trends in Cognitive Sciences* 14 (11): 506–15. https://doi.org/10.1016/j.tics.2010.09.001.

Carmona, J, J Suarez, and J Ochoa. 2017. "Brain Functional Connectivity in Parkinson's Disease - EEG Resting Analysis." In *VII Latin American Congress on Biomedical Engineering CLAIB 2016, Bucaramanga, Santander, Colombia, October 26th-28th, 2016*, 185–88.

Chen, X. L., Y. Y. Xiong, G. L. Xu, and X. F. Liu. 2012. "Deep Brain Stimulation." *Interventional Neurology* 1 (3–4): 200–2012. https://doi.org/10.1016/B978-0-12-385157-4.00740-5.

Cohen, Michael X. 2015. "Effects of Time Lag and Frequency Matching on Phase-Based Connectivity." *Journal of Neuroscience Methods* 250: 137–46. https://doi.org/10.1016/j.jneumeth.2014.09.005.

Cole, Michael W., Jeremy R. Reynolds, Jonathan D. Power, Grega Repovs, Alan Anticevic, and Todd S. Braver. 2013. "Multi-Task Connectivity Reveals Flexible Hubs for Adaptive Task Control." *Nature Neuroscience* 16 (9): 1348–55. https://doi.org/10.1038/nn.3470.

Costa, Luciano da F., Francisco A. Rodrigues, Gonzalo Travieso, and P. R. Villas Boas. 2007. "Characterization of Complex Networks: A Survey of Measurements" 56 (February 2007): 167–242. https://doi.org/10.1080/00018730601170527.

Cover, Thomas M, and Joy A Thomas. 2012. *Elements of Information Theory*. John Wiley & Sons.

Craddock, R. Cameron, Saad Jbabdi, Chao Gan Yan, Joshua T. Vogelstein, F. Xavier Castellanos, Adriana Di Martino, Clare Kelly, Keith Heberlein, Stan Colcombe, and Michael P. Milham. 2013. "Imaging Human Connectomes at the Macroscale." *Nature Methods* 10 (6): 524–39. https://doi.org/10.1038/nmeth.2482.

Davison, Elizabeth N., Kimberly J. Schlesinger, Danielle S. Bassett, Mary Ellen Lynall, Michael B. Miller, Scott T. Grafton, and Jean M. Carlson. 2015. "Brain Network Adaptability across Task States." *PLoS Computational Biology* 11 (1). https://doi.org/10.1371/journal.pcbi.1004029.

Deco, Gustavo, Viktor K. Jirsa, and Anthony R. McIntosh. 2011. "Emerging Concepts for the Dynamical Organization of Resting-State Activity in the Brain." *Nature Reviews Neuroscience* 12 (1): 43–56. https://doi.org/10.1038/nrn2961.

Diessen, Eric Van, Sander J.H. Diederen, Kees P.J. Braun, Floor E. Jansen, and Cornelis J. Stam. 2013. "Functional and Structural Brain Networks in Epilepsy: What Have We Learned?" *Epilepsia* 54 (11): 1855–65. https://doi.org/10.1111/epi.12350.

Domenico, Manlio De, Shuntaro Sasai, and Alex Arenas. 2016. "Mapping Multiplex Hubs in Human Functional Brain Networks." *Frontiers in Neuroscience* 10 (JUL): 1–14. https://doi.org/10.3389/fnins.2016.00326.

Domenico, Manlio De, Albert Solé-Ribalta, Emanuele Cozzo, Mikko Kivelä, Yamir Moreno, Mason A. Porter, Sergio Gómez, and Alex Arenas. 2014. "Mathematical Formulation of Multilayer Networks." *Physical Review X* 3 (4): 1–15. https://doi.org/10.1103/PhysRevX.3.041022.

Doron, K. W., D. S. Bassett, and M. S. Gazzaniga. 2012. "Dynamic Network Structure of Interhemispheric Coordination." *Proceedings of the National Academy of Sciences* 109 (46): 18661–68. https://doi.org/10.1073/pnas.1216402109.

Douw, Linda, Edwin van Dellen, Marjolein de Groot, Jan J. Heimans, Martin Klein, Cornelis J. Stam, and Jaap C. Reijneveld. 2010. "Epilepsy Is Related to Theta Band Brain Connectivity and Network Topology in Brain Tumor Patients." *BMC Neuroscience* 11. https://doi.org/10.1186/1471-2202-11-103.

Eguíluz, Victor M., Dante R. Chialvo, Guillermo A. Cecchi, Marwan Baliki, and A. Vania Apkarian. 2005. "Scale-Free Brain Functional Networks." *Physical Review Letters* 94 (1): 1–4. https://doi.org/10.1103/PhysRevLett.94.018102.

Engel, Andreas K., and Pascal Fries. 2016. *Neuronal Oscillations, Coherence, and Consciousness*. *The Neurology of Conciousness*. Elsevier Ltd. https://doi.org/10.1016/B978-0-12-800948-2.00003-0.

Engel, Andreas K., Christian K.E. Moll, Itzhak Fried, and George A. Ojemann. 2005. "Invasive Recordings from the Human Brain: Clinical Insights and Beyond." *Nature Reviews Neuroscience* 6 (1): 35–47. https://doi.org/10.1038/nrn1585.

Fornito, Alex, Andrew Zalesky, and Michael Breakspear. 2013. "Graph Analysis of the Human Connectome: Promise, Progress, and Pitfalls." *NeuroImage* 80: 426–44.

https://doi.org/10.1016/j.neuroimage.2013.04.087.

Fornito, Alex, Andrew Zalesky, Christos Pantelis, and Edward T. Bullmore. 2012. "Schizophrenia, Neuroimaging and Connectomics." *NeuroImage* 62 (4): 2296–2314. https://doi.org/10.1016/j.neuroimage.2011.12.090.

Friston, Karl J. 2011. "Functional and Effective Connectivity: A Review." *Brain Connectivity* 1 (1): 13–36. https://doi.org/10.1089/brain.2011.0008.

Friston, Karl J., Vladimir Litvak, Ashwini Oswal, Adeel Razi, Klaas E. Stephan, Bernadette C.M. Van Wijk, Gabriel Ziegler, and Peter Zeidman. 2016. "Bayesian Model Reduction and Empirical Bayes for Group (DCM) Studies." *NeuroImage* 128: 413–31. https://doi.org/10.1016/j.neuroimage.2015.11.015.

Ganella, E. P., C. Seguin, C. F. Bartholomeusz, S. Whittle, C. Bousman, C. M. Wannan, and A. Zalesky. 2018. "Default-Mode Ntwork Activity Distinguishes Alzheimer's Disease from Healthy Aging: Evidence from Functional MR." *Schizophrenia Research* 193: 284–92. https://doi.org/10.1016/j.schres.2017.07.014.

Garcés, Pilar, Ernesto Pereda, Juan A. Hernández-Tamames, Francisco Del-Pozo, Fernando Maestú, and José Ángel Pineda-Pardo. 2016. "Multimodal Description of Whole Brain Connectivity: A Comparison of Resting State MEG, FMRI, and DWI." *Human Brain Mapping* 37 (1): 20–34. https://doi.org/10.1002/hbm.22995.

Granger, C.W.J. 1969. "Investigating Causal Relations by Econometric Models and Cross-Spectral Methods Author(S):" *Ecometrica* 37 (Aug.,1969): 424–38. http://ir.obihiro.ac.jp/dspace/handle/10322/3933.

Grech, Roberta, Tracey Cassar, Joseph Muscat, Kenneth P. Camilleri, Simon G. Fabri, Michalis Zervakis, Petros Xanthopoulos, Vangelis Sakkalis, and Bart Vanrumste. 2008. "Review on Solving the Inverse Problem in EEG Source Analysis." *Journal of NeuroEngineering and Rehabilitation* 5. https://doi.org/10.1186/1743-0003-5-25.

Hallquist, M. N., and F. G. Hillary. 2018. "We Thank Zach Ceneviva, Allen Csuk, Richard Garcia, Melanie Glatz, and Riddhi Patel for Their Work Collecting, Organizing, and Coding References for the Literature Review and Manuscript." *Network Neuroscience*.

Haufe, Stefan, Vadim V. Nikulin, Klaus Robert Müller, and Guido Nolte. 2013. "A Critical Assessment of Connectivity Measures for EEG Data: A Simulation Study." *NeuroImage* 64 (1): 120–33. https://doi.org/10.1016/j.neuroimage.2012.09.036.

He, Biyu J., John M. Zempel, Abraham Z. Snyder, and Marcus E. Raichle. 2010. "The Temporal Structures and Functional Significance of Scale-Free Brain Activity." *Neuron* 66 (3): 353–69. https://doi.org/10.1016/j.neuron.2010.04.020.

Heuvel, M. P. van den, C. J. Stam, M. Boersma, and H. E. Hulshoff Pol. 2008. "Small-World and Scale-Free Organization of Voxel-Based Resting-State Functional Connectivity in the Human Brain." *NeuroImage* 43 (3): 528–39. https://doi.org/10.1016/j.neuroimage.2008.08.010.

Heuvel, Martijn P. Van Den, Olaf Sporns, Guusje Collin, Thomas Scheewe, René C.W. Mandl, Wiepke Cahn, Joaquín Goni, Hilleke E.Hulshoff Pol, and René S. Kahn. 2013. "Abnormal Rich Club Organization and Functional Brain Dynamics in Schizophrenia." *JAMA Psychiatry* 70 (8): 783–92. https://doi.org/10.1001/jamapsychiatry.2013.1328.

Hogan, R. Edward. 2018. "Malformations of Cortical Development: A Structural and Functional MRI Perspective." *Epilepsy Currents* 18 (2): 92–94. https://doi.org/10.5698/1535-7597.18.2.92.

Honey, Christopher J., Jean Philippe Thivierge, and Olaf Sporns. 2010. "Can Structure Predict Function in the Human Brain?" *NeuroImage* 52 (3): 766–76. https://doi.org/10.1016/j.neuroimage.2010.01.071.

Honey, Christopher J, Rolf Kö Tter †, Michael Breakspear, and Olaf Sporns. 2007. "Network Structure of Cerebral Cortex Shapes Functional Connectivity on Multiple Time Scales."

*Proceedings of the National Academy of Sciences* 104 (24): 10240–45. https://doi.org/10.1073/pnas.98.2.676.

Horn, Andreas, Dirk Ostwald, Marco Reisert, and Felix Blankenburg. 2014. "The Structural-Functional Connectome and the Default Mode Network of the Human Brain." *NeuroImage* 102 (P1): 142–51. https://doi.org/10.1016/j.neuroimage.2013.09.069.

Huang, Haiqing, Jared Tanner, Hari Parvataneni, Mark Rice, Ann Horgas, Mingzhou Ding, and Catherine Price. 2018. "Impact of Total Knee Arthroplasty with General Anesthesia on Brain Networks: Cognitive Efficiency and Ventricular Volume Predict Functional Connectivity Decline in Older Adults." *Journal of Alzheimer's Disease* 62 (1): 319–33. https://doi.org/10.3233/JAD-170496.

Huang, Weiyu, Thomas A W Bolton, John D Medaglia, Danielle S Bassett, Alejandro Ribeiro, and Dimitri Van De Ville. 2018. "Graph Signal Processing of Human Brain Imaging Data" 1: 980–84. https://doi.org/10.1021/acsnano.6b05674.

Humphries, Mark D. 2017. "Dynamical Networks: Finding, Measuring, and Tracking Neural Population Activity Using Network Science." *Network Neuroscience* 1 (4): 324–38. https://doi.org/10.1162/NETN_a_00020.

James, Ryan G., Nix Barnett, and James P. Crutchfield. 2016. "Information Flows? A Critique of Transfer Entropies." *Physical Review Letters* 116 (23): 1–5. https://doi.org/10.1103/PhysRevLett.116.238701.

Kabbara, A., H. Eid, W. El Falou, M. Khalil, F. Wendling, and M. Hassan. 2018. "Reduced Integration and Improved Segregation of Functional Brain Networks in Alzheimer's Disease." *Journal of Neural Engineering* 15 (2). https://doi.org/10.1088/1741-2552/aaaa76.

Kaiser, Marcus. 2011. "A Tutorial in Connectome Analysis: Topological and Spatial Features of Brain Networks." *NeuroImage* 57 (3): 892–907. https://doi.org/10.1016/j.neuroimage.2011.05.025.

Kelley, D. J., M. Farhoud, M. E. Meyerand, D. L. Nelson, L. F. Ramirez, R. J. Dempsey, ..., and R. J. Davidson. 2007. "Creating Physical 3D Stereolithograph Models of Brain and Skull." *PLoS ONE* 2 (10): e1119. https://doi.org/10.1371/journal.pone.0001119.

Khadem, Ali, Gholam Ali Hossein-Zadeh, and Anahita Khorrami. 2016. "Long-Range Reduced Predictive Information Transfers of Autistic Youths in EEG Sensor-Space During Face Processing." *Brain Topography* 29 (2): 283–95. https://doi.org/10.1007/s10548-015-0452-4.

Kivelä, Mikko, Alex Arenas, Marc Barthelemy, James P. Gleeson, Yamir Moreno, and Mason A. Porter. 2014. "Multilayer Networks." *Journal of Complex Networks* 2 (3): 203–71. https://doi.org/10.1093/comnet/cnu016.

Kuznetsov, Y. A. 2013. *Elements of Applied Bifurcation Theory*. Edited by Springer Science & Business Media. Vol. 112.

Lal, Thomas Navin, Thilo Hinterberger, Guido Widman, Michael Schr, Jeremy Hill, Wolfgang Rosenstiel, Christian E Elger, and Bernhard Sch. 2005. "Methods Towards Invasive Human Brain Computer Interfaces." *Advances in Neural Information Processing Systems* 17: 737–44. https://doi.org/10.1.1.64.8486.

Lee, Kyu Min, Byungjoon Min, and Kwang Il Goh. 2015. "Towards Real-World Complexity: An Introduction to Multiplex Networks." *European Physical Journal B* 88 (2). https://doi.org/10.1140/epjb/e2015-50742-1.

Lee Rodgers, J., and W. A. Nicewander. 1988. "Thirteen Ways to Look at the Correlation Coefficient." *The American Statistician* 42 (1): 59–66.

Li, Kaiming, Lei Guo, Jingxin Nie, Gang Li, and Tianming Liu. 2009. "Review of Methods for Functional Brain Connectivity Detection Using FMRI." *Computerized Medical Imaging and Graphics* 33 (2): 131–39. https://doi.org/10.1016/j.compmedimag.2008.10.011.

Lindner, Michael, Raul Vicente, Viola Priesemann, and Michael Wibral. 2011. "TRENTOOL: A Matlab Open Source Toolbox to Analyse Information Flow in Time Series Data with Transfer Entropy." *BMC Neuroscience* 12. https://doi.org/10.1186/1471-2202-12-119.

Lynall, M.-E., D. S. Bassett, R. Kerwin, P. J. McKenna, M. Kitzbichler, U. Muller, and E. Bullmore. 2010. "Functional Connectivity and Brain Networks in Schizophrenia." *Journal of Neuroscience* 30 (28): 9477–87. https://doi.org/10.1523/JNEUROSCI.0333-10.2010.

Ma, Sai, Vince D. Calhoun, Ronald Phlypo, and Tülay Adali. 2014. "Dynamic Changes of Spatial Functional Network Connectivity in Healthy Individuals and Schizophrenia Patients Using Independent Vector Analysis." *NeuroImage* 90: 196–206. https://doi.org/10.1016/j.neuroimage.2013.12.063.

MacKay, J. C. 2003. "Information Theory, Inference, and Learning Algorithms." *Cambridge University Press*. https://doi.org/10.1016/S0020-7063(14)00055-7.

Maris, Eric, Pascal Fries, and Freek van Ede. 2016. "Diverse Phase Relations among Neuronal Rhythms and Their Potential Function." *Trends in Neurosciences* 39 (2): 86–99. https://doi.org/10.1016/j.tins.2015.12.004.

Marreiros, Goreti, Ricardo Santos, Carlos Ramos, and José Neves. 2010. "Context-Aware Emotion-Based Model for Group Decision Making." *IEEE Intelligent Systems* 25 (2): 31–39. https://doi.org/10.1109/MIS.2010.46.

Meskaldji, Djalel Eddine, Lana Vasung, David Romascano, Jean Philippe Thiran, Patric Hagmann, Stephan Morgenthaler, and Dimitri Van De Ville. 2015. "Improved Statistical Evaluation of Group Differences in Connectomes by Screening-Filtering Strategy with Application to Study Maturation of Brain Connections between Childhood and Adolescence." *NeuroImage* 108: 251–64. https://doi.org/10.1016/j.neuroimage.2014.11.059.

Milo, R., S. Shen-Orr, S. Itkovitz, N. Kashtan, D. Chlovskii, and U. Alon. 2002. "Network Motifs : Simple Building Blocks of Complex Networks Author ( s ): R . Milo , S . Shen-Orr , S . Itzkovitz , N . Kashtan , D . Chklovskii and U . Alon Published by : American Association for the Advancement of Science Stable URL : Http://Www.Jstor." *Science* 298 (5594): 824–27.

Navlakha, Saket, Rajeev Rastogi, and Nisheeth Shrivastava. 2008. "Graph Summarization with Bounded Error." *Proceedings of the 2008 ACM SIGMOD International Conference on Management of Data - SIGMOD '08*, 419. https://doi.org/10.1145/1376616.1376661.

Newman, M. E. J. 2003. "The Structure and Function of Complex Networks." *SIAM Reveiw* 45 (2): 167–256. https://doi.org/10.1137/S003614450342480.

Nolte, Guido, Andreas Ziehe, Nicole Kramer, Florin Popescu, and Klaus-Robert Muller. 2008. "Comparison of Granger Causality and Phase Slope Index." *Nips* 6: 267–76.

Ouma, Wilberforce Zachary, Katja Pogacar, and Erich Grotewold. 2018. "Topological and Statistical Analyses of Gene Regulatory Networks Reveal Unifying yet Quantitatively Different Emergent Properties." *PLoS Computational Biology* 14 (4): 1–17. https://doi.org/10.1371/journal.pcbi.1006098.

Papo, David, Massimiliano Zanin, and Javier M. Buldú. 2014. "Reconstructing Functional Brain Networks: Have We Got the Basics Right?" *Frontiers in Human Neuroscience* 8 (February): 8–11. https://doi.org/10.3389/fnhum.2014.00107.

Park, Hae Jeong, and Karl Friston. 2013. "Structural and Functional Brain Networks: From Connections to Cognition." *Science* 342 (6158). https://doi.org/10.1126/science.1238411.

Pascual-Marqui, R. D. 1999. "Review of Methods for Solving the EEG Inverse Problem." *International Journal of Bioelectromagnetism* 1 (1): 75–86. https://doi.org/citeulike-article-id:5020586.

Pesaran, Bijan, Martin Vinck, Gaute T. Einevoll, Anton Sirota, Pascal Fries, Markus Siegel, Wilson Truccolo, Charles E. Schroeder, and Ramesh Srinivasan. 2018. "Investigating

Large-Scale Brain Dynamics Using Field Potential Recordings: Analysis and Interpretation." *Nature Neuroscience*, 1–17. https://doi.org/10.1038/s41593-018-0171-8.

Pizzagalli, D. A. 2007. "Electroencephalography and High-Density Electrophysiological Source Localization." In *Handbook of Psychophysiology*, 3:56–84. https://doi.org/10.1109/TNSRE.2007.903919.

Polani, D. 2013. "Kullback-Leibler Divergence." *Springer* 15: 1087–88. https://doi.org/10.1016/0378-4754(88)90061-4.

Poli, D., Pastore, V. P. Massobrio, P. 2015. "Functional Connectivity in in Vitro Neuronal Assemblies." *Frontiers in Neural Circuits* 9 (October): 57. https://doi.org/10.3389/fncir.2015.00057.

Rajula, Hema Sekhar Reddy, Matteo Mauri, and Vassilios Fanos. 2018. "Scale-Free Networks in Metabolomics." *Bioinformation* 14 (03): 140–44. https://doi.org/10.6026/97320630014140.

Razi, A., and K. J Friston. 2016. "The Connected Brain." *IEEE Signal Processing Magazine* 33 (3): 14–35.

Rissman, Jesse, Adam Gazzaley, and Mark D'Esposito. 2004. "Measuring Functional Connectivity during Distinct Stages of a Cognitive Task." *NeuroImage* 23 (2): 752–63. https://doi.org/10.1016/j.neuroimage.2004.06.035.

Rubinov, Mikail, and Olaf Sporns. 2010. "Complex Network Measures of Brain Connectivity: Uses and Interpretations." *NeuroImage* 52 (3): 1059–69. https://doi.org/10.1016/j.neuroimage.2009.10.003.

Schreiber, Thomas. 2006. "Measuring Information Transfer." *Physical Review Letters* 85 (2): 461. https://doi.org/10.1103/PhysRevLett.85.461.

Schweder, Tore. 1970. "Composable Markov Processes." *Journal of Applied Probability* 7 (Aug., 1970): 400–410. https://doi.org/10.1111/j.1365-2672.2007.03484.x.

Shaw, John C. 1984. "Correlation and Coherence Analysis of the EEG: A Selective Tutorial Review." *International Journal of Psychophysiology* 1 (3): 255–66. https://doi.org/10.1016/0167-8760(84)90045-X.

Shine, James M., Matthew J. Aburn, Michael Breakspear, and Russell A. Poldrack. 2018. "The Modulation of Neural Gain Facilitates a Transition between Functional Segregation and Integration in the Brain." *ELife* 7: 1–16. https://doi.org/10.7554/eLife.31130.

Shlens, Jonathon. 2014. "Notes on Kullback-Leibler Divergence and Likelihood." *ArXiv*, 1–4. https://doi.org/10.1108/JKM-06-2014-0253.

Sifuzzaman, M., M. R. Islam, and M. Z. Ali. 2009. "Application of Wavelet Transform and Its Advantages Compared To Fourier Transform."

Sporns, O. 2016. "Connectome Networks: From Cells to Systems." In *Micro-, Meso-and Macro-Connectomics of the Brain*, 107–27.

Sporns, Olaf. 2002. "Network Analysis, Complexity, and Brain Function." *Complexity* 8 (1): 56–60. https://doi.org/10.1002/cplx.10047.

———. 2011. "The Human Connectome: A Complex Network." *Annals of the New York Academy of Sciences* 1224 (1): 109–25. https://doi.org/10.1111/j.1749-6632.2010.05888.x.

———. 2013a. "Network Attributes for Segregation and Integration in the Human Brain." *Current Opinion in Neurobiology* 23 (2): 162–71. https://doi.org/10.1016/j.conb.2012.11.015.

———. 2013b. "The Human Connectome: Origins and Challenges." *NeuroImage* 80: 53–61. https://doi.org/10.1016/j.neuroimage.2013.03.023.

Sporns, Olaf, Dante R. Chialvo, Marcus Kaiser, and Claus C. Hilgetag. 2004. "Organization, Development and Function of Complex Brain Networks." *Trends in Cognitive Sciences* 8 (9): 418–25. https://doi.org/10.1016/j.tics.2004.07.008.

Sporns, Olaf, and Rolf Kötter. 2004. "Motifs in Brain Networks." *PLoS Biology* 2 (11). https://doi.org/10.1371/journal.pbio.0020369.

Sporns, Olaf, Giulio Tononi, and Rolf Kötter. 2005. "The Human Connectome: A Structural Description of the Human Brain." *PLoS Computational Biology* 1 (4): 0245–51. https://doi.org/10.1371/journal.pcbi.0010042.

Sporns, Olaf, and Jonathan D Zwi. 2004. "The Small World of the Cerebral Cortex." *Neuroinformatics* 2 (2): 145–62. https://doi.org/10.1385/NI:2:2:145.

Stam, C. J., B. F. Jones, G. Nolte, M. Breakspear, and Ph Scheltens. 2007. "Small-World Networks and Functional Connectivity in Alzheimer's Disease." *Cerebral Cortex* 17 (1): 92–99. https://doi.org/10.1093/cercor/bhj127.

Stam, C. J., and E. C.W. van Straaten. 2012. "The Organization of Physiological Brain Networks." *Clinical Neurophysiology* 123 (6): 1067–87. https://doi.org/10.1016/j.clinph.2012.01.011.

Stam, Cornelis J. 2014. "Modern Network Science of Neurological Disorders." *Nature Reviews Neuroscience* 15 (10): 683–95. https://doi.org/10.1038/nrn3801.

Stam, Cornelis J., and Jaap C. Reijneveld. 2007. "Graph Theoretical Analysis of Complex Networks in the Brain." *Nonlinear Biomedical Physics* 1: 1–19. https://doi.org/10.1186/1753-4631-1-3.

Stephan, Klaas E., Karl J. Friston, and Chris D. Frith. 2009. "Dysconnection in Schizophrenia: From Abnormal Synaptic Plasticity to Failures of Self-Monitoring." *Schizophrenia Bulletin* 35 (3): 509–27. https://doi.org/10.1093/schbul/sbn176.

Stephan, Klaas Enno, and Karl J. Friston. 2010. "Analyzing Effective Connectivity with Functional Magnetic Resonance Imaging." *Wiley Interdisciplinary Reviews: Cognitive Science* 1 (3): 446–59. https://doi.org/10.1002/wcs.58.

Strogatz, Steven H. 2001. "Exploring Complex Networks." *Nature* 410 (6825): 268–76. https://doi.org/10.1038/35065725.

Sun, Felice T., Lee M. Miller, and Mark D'Esposito. 2004. "Measuring Interregional Functional Connectivity Using Coherence and Partial Coherence Analyses of FMRI Data." *NeuroImage* 21 (2): 647–58. https://doi.org/10.1016/j.neuroimage.2003.09.056.

Supekar, Kaustubh, Vinod Menon, Daniel Rubin, Mark Musen, and Michael D. Greicius. 2008. "Network Analysis of Intrinsic Functional Brain Connectivity in Alzheimer's Disease." *PLoS Computational Biology* 4 (6). https://doi.org/10.1371/journal.pcbi.1000100.

Tecchio, Franca, Carlo Cottone, Camillo Porcaro, Andrea Cancelli, Vincenzo Di Lazzaro, and Giovanni Assenza. 2018. "Brain Functional Connectivity Changes After Transcranial Direct Current Stimulation in Epileptic Patients." *Frontiers in Neural Circuits* 12 (May): 1–7. https://doi.org/10.3389/fncir.2018.00044.

Thompson, William Hedley, Per Brantefors, and Peter Fransson. 2017. "From Static to Temporal Network Theory: Applications to Functional Brain Connectivity." *Network Neuroscience* 1 (2): 69–99. https://doi.org/10.1162/NETN_a_00011.

Vaiana, Michael, and Sarah Feldt Muldoon. 2018. "Multilayer Brain Networks." *Journal of Nonlinear Science*, no. September 2017: 1–23. https://doi.org/10.1007/s00332-017-9436-8.

Valdes-Sosa, Pedro A., Alard Roebroeck, Jean Daunizeau, and Karl Friston. 2011. "Effective Connectivity: Influence, Causality and Biophysical Modeling." *NeuroImage* 58 (2): 339–61. https://doi.org/10.1016/j.neuroimage.2011.03.058.

Vella, Danila, Italo Zoppis, Giancarlo Mauri, Pierluigi Mauri, and Dario Di Silvestre. 2017. "From Protein-Protein Interactions to Protein Co-Expression Networks: A New Perspective to Evaluate Large-Scale Proteomic Data." *Eurasip Journal on Bioinformatics and Systems Biology* 2017 (1). https://doi.org/10.1186/s13637-017-0059-z.

Vergara, Jorge R., and Pablo A. Estévez. 2014. "A Review of Feature Selection Methods Based

on Mutual Information." *Neural Computing and Applications* 24 (1): 175–86. https://doi.org/10.3762/bjnano.7.4.

Wang, X. F. 2002. "Complex Networks: Topology, Dynamics and Synchronization." *International Journal of Bifurcation and Chaos* 12 (05): 885–916.

Watts, Duncan J, and Steven H Strogatz. 1998. "Watts-1998-Collective Dynamics of 'small-World'" 393 (June): 440–42. https://doi.org/Doi 10.1038/30918.

Wiener, N. 1956. "The Theory of Prediction. Modern Mathematics for Engineers." *New York*, 165–90.

Wig, Gagan S., Bradley L. Schlaggar, and Steven E. Petersen. 2011. "Concepts and Principles in the Analysis of Brain Networks." *Annals of the New York Academy of Sciences* 1224 (1): 126–46. https://doi.org/10.1111/j.1749-6632.2010.05947.x.

Williams, Nitin, and Richard N. Henson. 2018. "Recent Advances in Functional Neuroimaging Analysis for Cognitive Neuroscience." *Brain and Neuroscience Advances* 2: 239821281775272. https://doi.org/10.1177/2398212817752727.

Wilson, Robin J. 1979. *Introduction to Graph Theory*. Pearson Education India.

Yoneki, Eiko, Pan Hui, and Jon Crowcroft. 2008. "Distinct Types of Hubs in Human Dynamic Networks." *Proceedings of the 1st Workshop on Social Network Systems - SocialNets '08*, 7–12. https://doi.org/10.1145/1435497.1435499.

Yu, Rongjun, Yi Ling Chien, Hsiao Lan Sharon Wang, Chih Min Liu, Chen Chung Liu, Tzung Jeng Hwang, Ming H. Hsieh, Hai Gwo Hwu, and Wen Yih Isaac Tseng. 2014. "Frequency-Specific Alternations in the Amplitude of Low-Frequency Fluctuations in Schizophrenia." *Human Brain Mapping* 35 (2): 627–37. https://doi.org/10.1002/hbm.22203.

Zhang, Han, Yu Jin Zhang, Chun Ming Lu, Shuang Ye Ma, Yu Feng Zang, and Chao Zhe Zhu. 2010. "Functional Connectivity as Revealed by Independent Component Analysis of Resting-State FNIRS Measurements." *NeuroImage* 51 (3): 1150–61. https://doi.org/10.1016/j.neuroimage.2010.02.080.

Zhang, Zhiqiang, Wei Liao, Huafu Chen, Dante Mantini, Ju Rong Ding, Qiang Xu, Zhengge Wang, et al. 2011. "Altered Functional-Structural Coupling of Large-Scale Brain Networks in Idiopathic Generalized Epilepsy." *Brain* 134 (10): 2912–28. https://doi.org/10.1093/brain/awr223.